\documentclass[reprint,aps,prd,amssymb, amsmath, nofootinbib, onecolumn]{revtex4-2}

\usepackage{graphicx}

\usepackage{mathtools}
\usepackage[svgnames]{xcolor} 

\usepackage[colorlinks=True, linkcolor=SlateBlue,
            citecolor=SteelBlue,urlcolor=BlueViolet]{hyperref}
\usepackage[capitalise]{cleveref}
\usepackage{orcidlink}
\usepackage{extpfeil}
\usepackage[T1]{fontenc}

\usepackage{aas_macros}

\newcommand{\vect}[1]{\boldsymbol{#1}}
\newcommand{\dvect}[1]{\dot{\boldsymbol{#1}}}
\newcommand{\uvect}[1]{\vect{\hat{#1}}}

\newcommand{\la}{\langle}
\newcommand{\ra}{\rangle}

\begin{document}

\author{Hang Yu\,\orcidlink{0000-0002-6011-6190}}
\email{hang.yu2@montana.edu}

\affiliation{eXtreme Gravity Institute, Department of Physics, Montana State University,
Bozeman, MT 59717, USA}

\author{Shu Yan Lau\,\orcidlink{0000-0002-8239-0174}}
\affiliation{eXtreme Gravity Institute, Department of Physics, Montana State University,
Bozeman, MT 59717, USA}



\title{Effective-one-body model for coalescing binary neutron stars:\\
Incorporating tidal spin and enhanced radiation from dynamical tides}

\begin{abstract}
    Tidal interactions in a coalescing binary neutron star (BNS) or neutron star-black hole (NSBH) system driven by gravitational wave (GW) radiation contain precious information about physics both at extreme density and in the highly relativistic regime. 
    In the late inspiral stage, where the tidal effects are the strongest, finite-frequency or even dynamical corrections to the tidal response become significant.
    Many previous analyses model the finite-frequency correction through the effective Love number approach, which only accounts for the correction in the radial interaction but ignores the lag in the tidal bulge behind the companion due to the continuous orbital shrinkage. The lag provides a torque between the orbit and the tidal bulge, causing the star's spin to change over time. We dub the evolving component of the spin the tidal spin, whose dimensionless value can reach 0.03-0.4 depending on how rapidly the background star rotates. 
    We present a relativistic, effective-one-body (EOB) waveform model for BNSs and NSBHs that incorporates the tidal spin, particularly its back reaction to the orbit due to the Newtonian tidal torque and the relativistic orbital hang-up.
    Beyond the conservative dynamics specified by the EOB Hamiltonian, we also derive the corrections to the dissipative GW radiation due to finite-frequency effects to the first post-Newtonian order.
    Depending on the star's background spin, the phase error in the time-domain waveform due to ignoring the tidal spin ranges from 0.3 to 4 radians at the waveform's peak amplitude.
    Notably, the difference in the waveforms with and without the tidal spin remarkably resembles the difference between previous effective Love number models and numerical relativity simulations, underscoring the significance of tidal spin in the construction of faithful models. 
    Our model further extends the description of dynamics in the high-background spin regions of the parameter space that are yet to be covered by numerical simulations. 
\end{abstract}

\maketitle

\section{Introduction}

Gravitational wave (GW) observations of the tidal signatures in coalescing binary neutron star (BNS) and neutron star-black hole (NSBH) systems may reveal crucial information about the neutron star (NS) equation of state (EoS) and extreme gravity as successfully demonstrated with the event GW170817 \cite{GW170817, GW170817eos}. 

While the tidal effects are most accurately studied by numerical relativity (NR) simulations~\cite{Hotokezaka:15, Dietrich:18, Foucart:19}, analytical models describing them are also necessary as they provide a first-principle explanation of NR results. More importantly, when performing parameter estimation tasks, an analytical model constructed from physical principles enables accurate interpolation and even extrapolation of simulation results over the parameter space. In fact, the underlying dimensionality of the relevant parameter space is best understood through analytical models. For example, while under the adiabatic limit, a single tidal deformability parameter (or equivalently, the Love number) fully characterizes the tidal response~\cite{Lai:94a, Lai:94b, Flanagan:08, Damour:09, Hinderer:10, Bini:12, Damour:12, Bini:14, Bernuzzi:15, Nagar:18}, analyses have shown that the finite-frequency (FF) response of the tidal modes (especially the fundamental modes or f-modes) can also be important~\cite{Steinhoff:16, Hinderer:16, Steinhoff:21}. In this case, the natural frequency of the NS f-mode represents another degree of freedom that has a complicated dependence on multiple factors including the mass \cite{Ranjan:24}, deformability \cite{Chan:14, Pradhan:23}, spin rate \cite{Doneva:13, Kruger:20, Kruger:23}, and nonlinear hydrodynamics \cite{Yu:23a, Kwon:24}. Identifying the relevant factors is essential in understanding the range of validity and potential limitations of NR-calibrated semi-analytical approximations \cite{Dietrich:17, Kawaguchi:18, Dietrich:19, Gamba:23b, Abac:24}. 

When constructing the analytical models, a commonly adopted treatment is the effective Love number approach \cite{Steinhoff:16, Hinderer:16, Schmidt:19, Steinhoff:21, Andersson:20, Andersson:21, Passamonti:22, Gamba:23, Kuan:23, Mandal:23, Pnigouras:24, Pitre:24, Pradhan:24}. In this framework, the constant tidal Love number is replaced with an effective, frequency-dependent one that captures the amplification in the tidal multiples due to f-mode resonance. The tidal bulge, in this picture, still points instantaneously toward the companion as in the adiabatic limit and the interaction remains in the radial direction. 

However, when the adiabatic assumption is relaxed, the tide also carries angular momentum (dubbed ``tidal spin'' in this work) that interacts with the orbit through a tidal torque~\cite{Friedman:78, Lai:94c, Schenk:02, Ma:20, Yu:24a} and evolves with the orbit (as opposed to the background spin that remains constant). Such a torque exists because the continuous orbital decay driven by GW causes the tidal bulge to lag behind the companion. Indeed, the orbital decay can be effectively viewed as a damping term on the tidal modes~\cite{Lai:94c, Yu:24a} and is likely to dominate over fluid dissipation in the late inspiral stage \cite{Weinberg:13, Venumadhav:14, Weinberg:16, Essick:16, GW170817pg, Arras:19, AbhishekHegade:24}. As shown by earlier analyses~\cite{Ma:20, Yu:24a}, this torque is missing in the effective Love number approach, yet it can be significant especially if the NS spins rapidly with the spin axis anti-aligned with the orbit, which allows the NS f-modes to be resonantly excited during the inspiral.
\cite{Ma:20, Yu:24a} are conducted under the Newtonian dynamics and therefore cannot be directly used in data analysis. They also missed post-Newtonian (PN) spin-orbit and spin-spin interactions \cite{Campanelli:06} due to the evolving tidal spin excited by the tidal torque.
In this work, we extend the analysis to incorporate general relativistic (GR) orbital dynamics via the effective-one-body (EOB) approach \cite{Buonanno:99} to produce waveform models ready to use in actual data analysis. 

\begin{figure}
    \centering
    \includegraphics[width=0.75\linewidth]{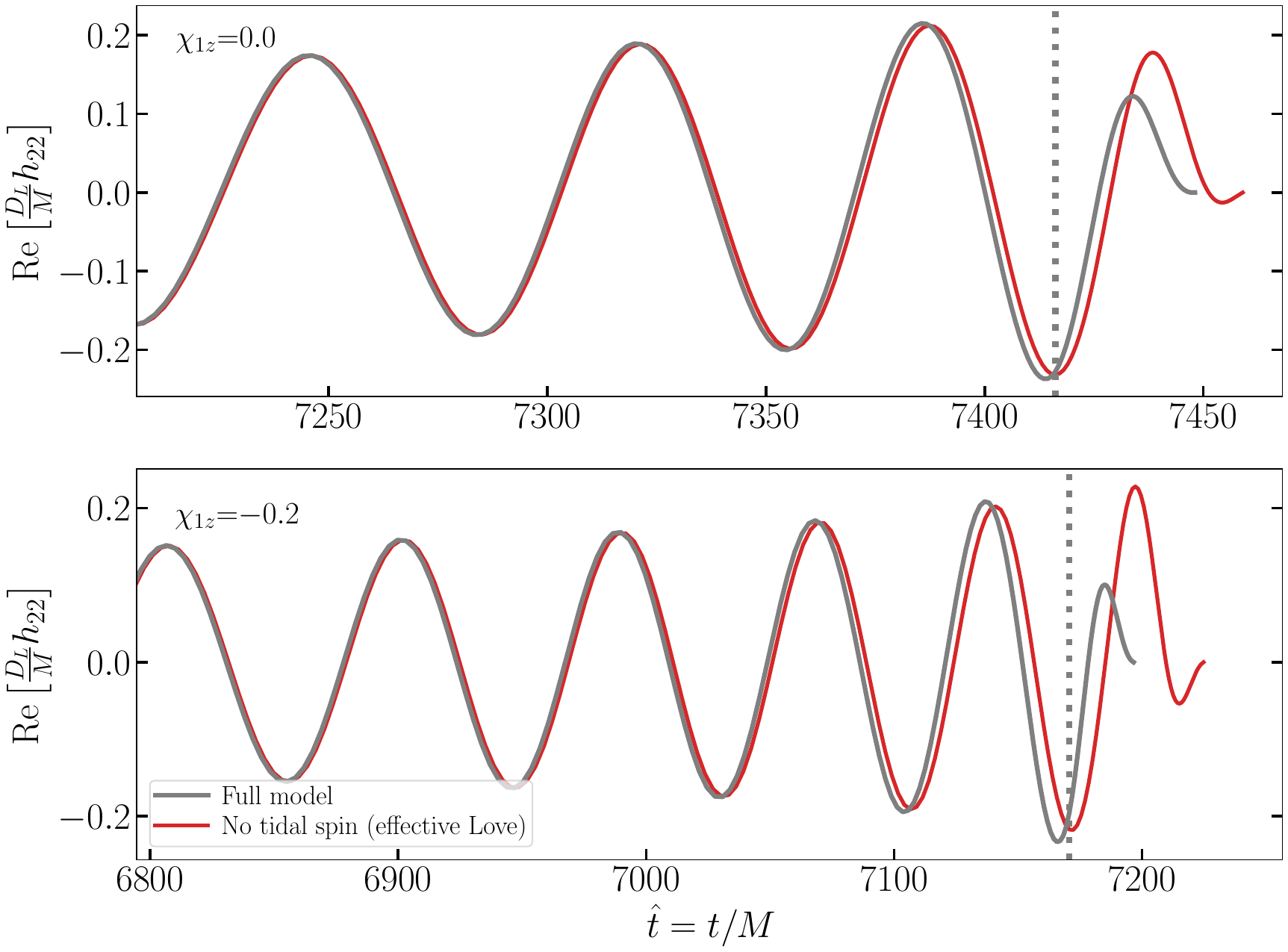}
    \caption{Normalized (2,2) GW mode $h_{22}$ of an equal mass BNS with masses $M_1=M_2=1.35\,M_\odot$ and aligned dimensionless background spin $\chi_{1z}=0$ (top) or $\chi_{1z}=-0.2$ (bottom). The other NS always has zero background spin.
    The vertical, dotted line marks the location where $|h_{22}|$ reaches its maximum.
    The gray curve is from our full tidal EOB model, while in the red curve, the tidal spin is zeroed when computing the back reaction on the orbit (which disables both the Newtonian torque and the post-Newtonian orbital hang-up due to the tidal spin-orbit coupling and reduces essentially to an effective Love number model).  
    The difference between the two sets of curves is remarkably similar to the difference between NR and the model of \cite{Steinhoff:21} based on the effective Love number approach (see their fig. 3). This underscores the significance of incorporating the tidal spin in producing faithful waveform models for BNSs. }
    \label{fig:h22_tidal_spin}
\end{figure}

An example GW waveform produced from our new model is presented in Fig. \ref{fig:h22_tidal_spin} for an equal-mass BNS system with the SLy EoS~\cite{Douchin:01} evolved to contact. The top panel assumes both NSs to have zero background spins. In the bottom panel, one NS has a background spin anti-aligned with the orbit corresponding to a dimensionless spin of $\chi_{1z}=-0.25$ with the $z$ axis marking the direction of the orbital angular momentum (AM). 
The waveforms are aligned at $\hat{t}=t/M=0$ with $M=M_1+M_2$ the total mass, corresponding to an orbital separation of $\hat{r} = r/M=18$, or a GW frequency of $f\simeq 300\,{\rm Hz}$.
The vertical dashed line marks the location where the amplitude of the (2, 2) GW mode reaches its peak. 
The gray curves are computed from the full model, and to get the red curves, we disable the tidal spins when computing the back reactions on the orbit (which essentially reduces to the effective Love number approach). While the details will be presented in the following sections, we highlight that the difference between the two sets of curves in Fig.~\ref{fig:h22_tidal_spin} is remarkably similar to the difference between NR simulations and the model of \cite{Steinhoff:21} (see the top two panels of their fig. 3). This indicates that the tidal spin included in our new model is an essential component when constructing faithful waveform models that match NR. 

Besides the tidal torque, another key improvement of our study is incorporating the FF tidal effects in the production of dissipative GW radiation. Previously, tidally induced GW modes were either approximated under the adiabatic limit or by a replacement of the Love number with its effective value \cite{Hinderer:16}. Neither gives the correct Newtonian mode resonance. Furthermore, the previous analysis did not properly separate the equilibrium (phase coherent with the orbit) and dynamical (varying at the mode's natural frequency and incoherent with the orbit) components of the tide when computing the radiation. 
In this work, we self-consistently include the FF effects in the tidally induced GW modes to the 1 PN order.

Throughout the paper, we use geometrical units $G=c=1$. We use Greek letters to denote spacetime indices that run over $\{0, 1, 2, 3\}$ and Latin letters to denote 3-dimensional spatial components. We adopt the convention of \cite{Yu:24a} where the tide is decomposed based on its phase evolution. In particular, the dynamical tide refers specifically to the component that varies at a mode's own natural frequency, and the equilibrium tide refers to the component whose phase follows the orbit. In this convention, the equilibrium tide includes FF corrections. The adiabatic tide is then used when a mode's natural frequency approaches infinity and the FF response reduces to unity.

The rest of the paper is organized as follows. We first introduce the conservative part of the dynamics specified by various Hamiltonians in Sec. \ref{sec:cons_dyn}. We start by discussing the Newtonian dynamics (Secs. \ref{sec:H_ns_N}-\ref{sec:tidal_spin}), which is sufficient to show the appearance of the tidal spin when FF effects are included. We also demonstrate explicitly that even when the full Hamiltonian in \cite{Hinderer:16} is used (instead of the effective Love number approximation), it does not give the correct Newtonian torque. After correcting for this, we utilize results from \cite{Hinderer:16, Steinhoff:16, Steinhoff:21} to incorporate PN effects in Sec. \ref{sec:H_PN} and resum it to the EOB form in Sec. \ref{sec:eob_hamiltonian}. The dissipative GW radiation including FF corrections in the tide is presented in Sec. \ref{sec:GW_rad}. Combining both conservative and dissipative parts, our final waveform is shown in Sec. \ref{sec:results}. Lastly, we conclude and discuss the limitations and future directions of our model in Sec. \ref{sec:discussion}.

\section{Conservative dynamics}
\label{sec:cons_dyn}

\subsection{Newtonian Hamiltonian in the frame corotating with the neutron star}
\label{sec:H_ns_N}

We start our discussion in a frame corotating with the tidally deformed NS (with mass $M_1$ and radius $R_1$). We abbreviate this as the ``NS frame'' and use the subscript ``ns'' to denote quantities evaluated in this frame when needed. The companion $M_2$ is treated as a point particle (pp) in the theoretical derivations. In most cases, tidal effects in $M_2$ can be obtained from the $M_1$ results by simply swapping the subscripts 1 and 2 (except for the tidal correction to the system's spin quadrupole and mass octupole where an additional minus sign is needed; see Sec.~\ref{sec:GW_rad}). 

For a perturbed fluid parcel with a Lagrangian displacement $\vect{\xi}_{\rm ns}$ at position $\vect{x}_1$ inside the NS, its Newtonian equation of motion, to the first order in the background spin $\Omega_1= |\vect{\Omega}_1|$, can be derived from the Lagrangian density (only for the tidal part as indicated by the subscript ``t''; see eq. A1 of \cite{Schenk:02} with a typo fixed; see also eq. (2.19) of Ref. \cite{Steinhoff:21})
\begin{equation}
    \mathcal{L}_{\rm t, ns} = \frac{1}{2} \dvect{\xi}_{\rm ns} \cdot \dvect{\xi}_{\rm ns} 
    - \vect{\xi}_{\rm ns} \cdot (\vect{\Omega}_1 \times \dvect{\xi}_{\rm ns}) 
    - \frac{1}{2} \vect{\xi}_{\rm ns} \cdot \vect{\mathcal{C}}\vect{\xi}_{\rm ns} + \vect{a}_{\rm ext,ns} \cdot \vect{\xi}_{\rm ns},
    \label{eq:Lagrangian_density_vs_xi}
\end{equation}
where the overdot denotes time derivative. The operator $\vect{\mathcal{C}}$ describes the internal restoring and $\vect{\Omega}_1$ is the spin of the background NS and in this work we restrict to the case where the spin is either aligned or anti-aligned with the orbital AM (i.e., along the $z$ axis). We further denote the external acceleration as $\vect{a}_{\rm ext,ns}$, which in our study is caused by the electric tidal potential,
\begin{equation}
    \vect{a}_{\rm ext,ns} = -(\nabla U)_{\rm ns} =  \sum_{l, m} \frac{M_2}{r^{l+1}} W_{lm} \nabla(x_1^l Y_{lm}) e^{- i m (\phi - \Omega_1 t)}, 
\end{equation}
where $U=-M_2/|\vect{r} - \vect{x}_1|$ is the interaction potential, $\vect{x}_1$ is a displacement vector of the perturbed fluid measured from the center of the NS $M_1$, and $r$ and $\phi$ are the orbital separation and phase. We use $Y_{lm}$ to denote spherical harmonics and $W_{lm}=4\pi Y_{lm}(\pi/2, 0)/(2l+1)$. 
The canonical momentum conjugate to $\vect{\xi}_{\rm ns}$ is 
\begin{equation}
    \vect{p}_{\rm ns}= \frac{\partial \mathcal{L}_{\rm t, ns}}{\partial \dot{\vect{\xi}}_{\rm ns}} = \dvect{\xi}_{\rm ns} + \vect{\Omega}_1 \times \vect{\xi}_{\rm ns}, 
\end{equation}
and the Hamiltonian density is given by 
\begin{align}
    \mathcal{H}_{\rm t, ns} &= \vect{p}_{\rm ns} \cdot \dvect{\xi}_{\rm ns} - \mathcal{L}_{\rm ns}  \nonumber \\
    &=\frac{1}{2}\dvect{\xi}_{\rm ns} \cdot \dvect{\xi}_{\rm ns} + \frac{1}{2} \vect{\xi}_{\rm ns} \cdot \vect{\mathcal{C}}\vect{\xi}_{\rm ns} - \vect{a}_{\rm ext,ns} \cdot \vect{\xi}_{\rm ns}, \label{eq:H_dxi_xi} \\
    &= \frac{1}{2} (\vect{p}_{\rm ns} - \vect{\Omega} \times \vect{\xi}_{\rm ns})^2 + \frac{1}{2} \vect{\xi}_{\rm ns} \cdot \vect{\mathcal{C}}\vect{\xi}_{\rm ns} - \vect{a}_{\rm ext,ns} \cdot \vect{\xi}_{\rm ns}.
\end{align}
We perform a \emph{phase} space expansion\footnote{We do not use a \emph{configuration} space expansion $\vect{\xi}=\sum_a q_a^{\rm (config)} \vect{\xi}_a^{(\rm config)}$ as did in \cite{Steinhoff:21} because when rotation is considered, the eigenfunctions $\xi_a^{\rm (config)}$ are not orthogonal for modes with different eigenfrequencies, and the equations of motion for $q_a^{(\rm config)}$ are not decoupled for different modes \cite{Schenk:02}. When only the f-modes are considered as in \cite{Steinhoff:21}, the issue of mode orthogonality can be ignored. Yet for a general treatment, the phase space expansion is preferred. } 
of the displacement field into eigenmodes~\cite{Schenk:02}
\begin{equation}
    \begin{bmatrix}
        \vect{\xi}_{\rm ns} (\vect{x}_1, t) \\
        \dot{\vect{\xi}}_{\rm ns}(\vect{x}_1, t)
    \end{bmatrix} 
    =\sum_a q_{a,{\rm ns}}(t)
    \begin{bmatrix}
        \vect{\xi}_a (\vect{x}_1) \\
        - i \omega_a {\vect{\xi}}_a (\vect{x}_1)
    \end{bmatrix} ,
    \label{eq:mode_decomp}
\end{equation}
where $\omega_a$ is the eigenfrequency of mode $a$ in the NS frame. In the expansion, a mode $a$ is labeled with four quantum numbers: its radial, polar, and azimuthal quantum numbers $(n_a, l_a, m_a)$ as well as the sign of its eigenfrequency $(s_a)$. 
The reality condition requires the amplitude and eigenfunction of a mode with $(m_a, s_a)$ are the complex conjugates of those of a mode with $(-m_a, -s_a)$ but the same $n_a$ and $l_a$ (same as the convention in \cite{Schenk:02}). That is, 
\begin{equation}
    q_{(n_a, l_a, m_a, s_a), {\rm ns}} = q_{(n_a, l_a, -m_a, -s_a), {\rm ns}}^\ast, \text{ and }
    \vect{\xi}_{(n_a, l_a, m_a, s_a), {\rm ns}} = \vect{\xi}_{(n_a, l_a, -m_a, -s_a), {\rm ns}}^\ast.
    \label{eq:real_cond}
\end{equation} 
Further, two modes $a$ and $b$ satisfy the orthogonality condition 
\begin{equation}
    (\omega_a + \omega_b) \la\vect{\xi}_a, \vect{\xi}_b \ra +  \la \vect{\xi}_a, 2i \vect{\Omega}_1 \times \vect{\xi}_b \ra = \epsilon_a \delta_{ab},
\end{equation}
where we keep the normalization general and introduce
\begin{align}
    \epsilon_a &= 2 \left\la\vect{\xi}_a, \omega_a \vect{\xi}_a + i \vect{\Omega}_1\times \vect{\xi}_a \right\ra = 2 \omega_{a0} \left\la \vect{\xi}_a, \vect{\xi}_a\right\ra, \\
    \text{with } &\la\vect{A}, \vect{B}\ra\equiv \int \vect{A}^\ast \cdot \vect{B} \rho(\vect{x}_1) d^3 x_1  ,
\end{align}
and $\omega_{a0}$ is the mode frequency if the background NS is non-spinning. 
We have used the fact that the shift of the NS-frame mode frequency satisfies~\cite{Christensen-Dalsgaard:98}
\begin{equation}
    \Delta \omega_a = \omega_a - \omega_{a0} = \frac{\left\la \vect{\xi}_a,  -i \vect{\Omega}_1 \times \vect{\xi}_a\right\ra}{\left\la \vect{\xi}_a, \vect{\xi}_a\right\ra} \equiv m_a C_a \Omega_1. 
    \label{eq:domega_a_coriolis}
\end{equation}
Note the frequency shift is proportional to $m_a\Omega_1$ to the leading order, and we have introduced a structural constant $C_a\simeq -1/l_a$ for Newtonian f-modes \cite{Yu:24a}. 
\cite{Steinhoff:21}, however, suggested that it is different for a relativistic star.\footnote{See their eqs. (5.4) and (5.7). Note that their $\omega_f$ is the f-mode frequency in the inertial frame, which is related to our NS-frame frequency $\omega_a$ through $\omega_f=\omega_a + m_a \Omega_1$. Further,  their $\bar{C}_{\rm SQ}$ is related to our $C_a$ through $\bar{C}_{\rm SQ}/I_1 = - 1 - C_a$ where $I_1$ is the NS's moment of inertia.} See also \cite{Kruger:20}. 
When acting on an eigenmode, the operator $\vect{\mathcal{C}}$ satisfies 
\begin{equation}
    \vect{\mathcal{C}}\vect{\xi}_a = \omega_a^2 \vect{\xi}_a + 2i \omega_a \vect{\Omega}_1 \times \vect{\xi}_a,
    \label{eq:C_operator}
\end{equation}
allowing us to write the Hamiltonian (obtained by integrating the density $\mathcal{H}_{\rm t, ns}$ over the NS) in terms of the mode amplitudes as~\cite{Flanagan:07} 
\begin{align}
    H_{\rm t, ns} = \sum_{a}^{+} \left(H_{a,{\rm ns}} + H_{a,{\rm int}}\right),
    \label{eq:H_NS_frame}
\end{align}
where the summation runs over modes with $s_a=+1$ and it is denoted by a ``$+$'' above the summation symbol. We define
\begin{align}
    H_{a, {\rm ns}} &= \epsilon_a \omega_a q_{a,{\rm ns}}^\ast q_{a,{\rm ns}}, \label{eq:H_a_ns}\\
    H_{a, {\rm int}} &= - \epsilon_a \omega_{a0} (u_{a, {\rm ns}} q_{a,{\rm ns}}^\ast + u_{a, {\rm ns}}^\ast q_{a,{\rm ns}}),\label{eq:H_a_int_ns}
\end{align}
where
\begin{equation}
    u_{a, {\rm ns}} = \frac{\la \vect{\xi}_a, \vect{a}_{\rm ext,ns} \ra}{\omega_{a0}\epsilon_a} = \left(\frac{M_1^2/R_1}{\omega_{a0} \epsilon_a}\right) W_{lm} I_a \left(\frac{M_2}{M_1}\right)\left(\frac{R_1}{r}\right)^{l+1} e^{- i m_a (\phi -\Omega_1 t) }. 
\end{equation}
with the tidal overlap integral $I_a$ given by
\begin{equation}
    I_a \equiv \frac{\la \vect{\xi}_a, \nabla (x_1^l Y_{lm})\ra}{M_1 R_1^l}.
    \label{eq:overlap}
\end{equation}
The reality condition in Eq.~(\ref{eq:real_cond}) indicates that $I_{m_a, s_a}=-I_{-m_a, -s_a}$ \cite{Yu:24a}. 
The Hamiltonian $H_{\rm t, ns}$ depends on time $t$ explicitly because $u_{a, {\rm ns}}$ contains $t$ explicitly.

The canonical displacements and momenta for the Hamiltonian in Eq.~(\ref{eq:H_NS_frame}) are $q_{a,{\rm ns}}$  and $ p_{a, {\rm ns}} = i \epsilon_a q_{a,{\rm ns}}^\ast$, respectively, where the modes are restricted to those with positive eigenfrequencies ($s_a=+$). In other words, the Poisson bracket is
\begin{equation}
    \left\{q_{a,{\rm ns}}, p_{a, {\rm ns}} \right\} = \left\{q_{a,{\rm ns}}, i \epsilon_a q_{b,{\rm ns}}^\ast \right\} = \delta_{ab}, \text{ with } \omega_{a,b}>0. 
\end{equation}
The equation of motion for each mode in the NS frame is thus
\begin{align}
    \dot{q}_{a, {\rm ns}} &= -\frac{i}{\epsilon_{a}} \frac{\partial H_{\rm t, ns}}{\partial q_{a, {\rm ns}}^{\ast}} = \left\{q_{a, {\rm ns}}, H_{\rm ns} \right\} 
    = - i \omega_{a} q_{a, {\rm ns}} + i \omega_{a0} u_{a, {\rm ns}}. 
\end{align}
The amplitudes of negative-frequency modes can be obtained from the reality requirement in Eq.~(\ref{eq:real_cond}). Throughout the analysis, we ignore damping of the f-mode (except for the effective damping induced by the orbital decay; see later in Eq. \ref{eq:b_a_eq_N_w_im}), as the expected dissipation rate due to shear viscosity \cite{Lai:94c}, Urca process \cite{Arras:19}, and GW radiation (due to the f-mode quadrupole coupling with itself; \cite{Yu:24a}) are all much smaller than $\omega_{a0}$.

\subsection{Changing reference frames}
\label{sec:frames}

To couple the mode dynamics with the orbital motion, it is convenient to transfer to the inertial frame (labeled with a subscript ``in''). Fast oscillations due to the orbital motion can be further eliminated by transferring to a frame corotating with the orbit so that the companion is fixed in the positive $x$-axis, which we will abbreviate as the ``orbit frame'' and label it with a subscript ``or''. At the Hamiltonian level, we adopt canonical transformations to switch between reference frames~\cite{Goldstein:02}. 

First, to go from the NS frame where each mode is described by $(q_{a, {\rm ns}},\ p_{a,{\rm ns}}{=}i\epsilon_a q_{a, {\rm ns}}^\ast)$ to the inertial frame with $(q_{a, {\rm in}},\ p_{a,{\rm in}})$, we adopt a (type-III) generator $G_{\rm ns-in} = {\rm g}_{\rm ns-in} + q_{a, {\rm ns}} p_{a, {\rm ns}}$ with
\begin{equation}
    {\rm g}_{\rm ns-in}(p_{a, {\rm ns}}, q_{a,{\rm in}}, t) = - p_{a, {\rm ns}} q_{a, {\rm in}} e^{i m_a \Omega_1 t},
\end{equation}
We then solve
\begin{align}
    &q_{a, {\rm ns}} = - \frac{\partial {\rm g}_{\rm ns-in}}{\partial p_{a,{\rm ns}}} = q_{a, {\rm in}} e^{i m_a \Omega_1 t},  \text{ or }  q_{a, {\rm in}} = q_{a, {\rm ns}} e^{-i m_a \Omega_1 t},\\
    &p_{a, {\rm in}} = - \frac{\partial {\rm g}_{\rm ns-in}}{\partial q_{a, {\rm in}}} = i \epsilon_a q_{a, {\rm in}}^\ast = i \epsilon_a q_{a, {\rm ns}}^\ast e^{i m_a \Omega_1 t},
\end{align}
which are both consistent with a Doppler shift of the mode's phase as expected,
and Hamiltonian transforms as
\begin{align}
    H_{\rm t, in}(q_{a, {\rm in}}, p_{a, {\rm in}}) &= H_{\rm t, ns}(q_{a, {\rm ns}}, p_{a, {\rm ns}}, t) + \frac{\partial {\rm g}_{\rm ns-in}(p_{a, {\rm ns}}, q_{a,{\rm in}}, t)}{\partial t} \nonumber \\
    &=\sum_a^+ \left(H_{a, {\rm ns}} + \Omega_1 S^{\rm}_a + H_{a, {\rm int}}\right),
    \label{eq:H_t_in}
\end{align}
where the mode ($H_{a, {\rm ns}}$) and interaction ($ H_{a, {\rm int}}$) pieces are the same as Eqs.~(\ref{eq:H_a_ns}) and (\ref{eq:H_a_int_ns}) except for that they are now evaluated in terms of $(q_{a, {\rm in}}, p_{a, {\rm in}})$ instead of $(q_{a, {\rm ns}}, p_{a, {\rm ns}})$, and we replace $u_{a, {\rm ns}}$ by $u_{a, {\rm in}} = u_{a, {\rm ns}}e^{- i \Omega_1 t}$. 
The additional piece is 
\begin{equation}
    S_{a} = \frac{1}{\Omega_1} \frac{\partial g_{\rm ns-in}}{\partial t} = \frac{m_a}{\omega_a} H_{a,{\rm ns}} = m_a \epsilon_a q_{a, {\rm in}} q_{a, {\rm in}}^\ast,
    \label{eq:S_a_can}
\end{equation}
which corresponds to the \emph{canonical} spin carried by a mode $a$, and $H_a\equiv(H_{a, {\rm ns}} + \Omega_1 S_{a})$ the mode's canonical energy in the inertial frame~\cite{Friedman:78}.\footnote{As noted in \cite{Friedman:78} and \cite{Schenk:02}, there are subtle differences between the canonical and physical spins of a mode. We provide more clarifications on this issue and show that using the canonical spin is appropriate in Appx.~\ref{appx:can_vs_phy_spin}. Because we ignore dissipation of the f-mode, the tidal spin will not be converted to the background in this study. } We also refer to $S_{a}$ as the ``tidal spin'' to be distinguished from the spin carried by the background star ($S_1=I_1\Omega_1=M_1^2 \chi_1$ with $I_1$ the moment of inertia). 
In particular, the background spin remains constant while the tidal spin carries the time-dependent evolution of the NS's AM in the canonical picture (as will be discussed in Fig. \ref{fig:chi1_t} later). 
As expected, the inertial frame Hamiltonian does not depend on time explicitly and is conserved in the absence of GW radiation. 
The equation of motion for a mode in the inertial frame is 
\begin{align}
    \dot{q}_{a, {\rm in}} &= -\frac{i}{\epsilon_{a}} \frac{\partial H_{\rm t, in}}{\partial q_{a, {\rm in}}^{\ast}} = \left\{q_{a, {\rm in}}, H_{\rm t, in} \right\} 
    = - i (\omega_{a} + m_a \Omega_1) q_{a, {\rm in}} + i \omega_{a0} u_{a, {\rm in}}. 
    \label{eq:dq_a_in}
\end{align}

The full Hamiltonian of the system is completed by adding to the tidal piece the pp contribution,
\begin{equation}
    H_{\rm in} (r, p_r, \phi, p_\phi, q_{a, {\rm in}}, p_{a, {\rm in}}) = H_{\rm t, in}(q_{a, {\rm in}}, p_{a, {\rm in}}) + H_{\rm pp}(r, p_r, \phi, p_\phi),
\end{equation}
where 
\begin{equation}
    H_{\rm pp} = \frac{p_r^2}{2 \mu} + \frac{p_\phi^2}{2 \mu r^2} - \frac{\mu M}{r}.
\end{equation}
In the above equation, $p_r$ and $p_\phi$ are the canonical momenta associated with $r$ and $\phi$, respectively. At the Newtonian order, they are given by $p_r=\mu \dot{r}$ and $p_\phi = \mu r^2 \dot{\phi}$. 

Going from the inertial frame $(q_{a, {\rm in}}, p_{a,{\rm in}})$ to the orbit frame proceeds in a similar manner. 
Because we will discuss dynamics extensively in both the inertial frame and the orbit frame, from this point onward, we will drop the ``in'' in the subscripts and use $(q_a, p_a)$ to denote specifically the mode amplitude and its conjugate momentum evaluated in the inertial frame. The corresponding quantities in the orbit frame will be denoted specifically with $(b_a, d_a)$. 
To remove the fast oscillation varying $m_a \phi$, we simultaneous transfer $(q_a, p_a, \phi, p_\phi)$ to $(b_a, d_a, Q_\phi, P_\phi)$ with a mixed generator $G_{\rm in-or} = {\rm g}_{\rm in-or} - Q_\phi P_\phi + q_a p_a$ (that is, type-II in $\phi$ and type-III in $q_a$) where
\begin{equation}
    {\rm g}_{\rm in-or}(p_a, b_a, \phi, P_\phi) = - p_a b_a e^{-i m_a \phi} + \phi P_\phi, 
    \label{eq:gen_in2or}
\end{equation}
satisfying 
\begin{align}
    &q_a = - \frac{\partial {\rm g}_{\rm in-or}}{\partial p_a}, 
    &d_a = -\frac{\partial {\rm g}_{\rm in-or}}{\partial b_a}, \\
    &Q_\phi = \frac{\partial {\rm g}_{\rm in-or}}{\partial P_\phi}, 
    &p_\phi = -\frac{\partial {\rm g}_{\rm in-or}}{\partial \phi}, 
\end{align}
which leads to
\begin{align}
    &b_a = q_a e^{i m_a \phi}, 
    &d_a= p_a e^{- i m_a \phi} =i \epsilon_a b_a^\ast, \\
    &Q_\phi = \phi, 
    &P_\phi = p_\phi + \sum_a^+ S_a.
\end{align}
The mode transfers again following a Doppler shift. For the orbit, the new canonical momentum conjugate to $\phi$ now becomes the ``\emph{total}'' AM of the system, including the pp orbital AM $p_\phi$ as well as the tidal spin $S_a$ (note we consider here non-precessing binaries so the background spin vectors $\vect{S}_{i} = I_{i}\vect{\Omega}_{i} = M_i^2\vect{\chi}_i$ with $i=1,2$ remain constant in the canonical picture and can be dropped from the total AM).  

In terms of the new canonical variables, the total Hamiltonian in the orbital frame reads
\begin{align}
    H_{\rm or}^{\rm (N)} &= \frac{p_r^2}{2 \mu} + \frac{\left(P_\phi - \sum_a^+ S_a \right)^2}{2\mu r^2} - \frac{\mu M}{r} + \sum_a^+ \left(H_a + H_{a,{\rm int}} \right) \nonumber \\
    &\simeq  \frac{p_r^2}{2 \mu} + \frac{P_\phi^2}{2 \mu r^2} - \frac{\mu M}{r} + \sum_a^+ \left[H_{a} - \frac{P_\phi}{\mu r^2 } S_a + H_{a,{\rm int}} \right].
    \label{eq:H_or_N}
\end{align}
where in the second line we have separated out the linear in $P_\phi$ term (cf. eq. 5.23 of \cite{Steinhoff:16}) and dropped the $S_a^2$ terms. 
In terms of $(b_a, d_a = i\epsilon_a b_a^\ast)$, we define
\begin{align}
    H_{\rm mode}\equiv&\sum_a^+ H_a \equiv \sum_a^+ \left(H_{a, {\rm ns}} + \Omega_1 S_a\right) = \sum_a^+ \left[\epsilon_a (\omega_a + m_a \Omega_1) b_a b_a^\ast\right] = \sum_a^+ \left[- i (\omega_a + m_a \Omega_1) b_a d_a\right], \label{eq:H_mode}\\
    S_{1z, {\rm mode}}\equiv & \sum_a^+ S_a = \sum_a^+ \frac{m_a}{\omega_a} H_{a, {\rm ns}} =\sum_a^+ m_a \epsilon_a b_a b_a^\ast = \sum_a^+ \left(-i m_a b_a d_a\right), \label{eq:S_a_vs_H_a}\\
    H_{{\rm int}, lm}\equiv&\sum_{(l_a, m_a){=}(l, m)}^+ H_{a, {\rm int}} = \sum_{(l_a, m_a){=}(l, m)}^+\left[-\epsilon_a \omega_{a0}v_a\left(b_a + b_a^\ast\right)\right] = \sum_{(l_a, m_a){=}(l, m)}^+\left[\omega_{a0} v_a (- \epsilon_a b_a + i d_a )\right], \\
    \text{with }  v_a =& u_{a, {\rm in}} e^{i m_a \phi} = \left(\frac{M_1^2/R_1}{\omega_{a0} \epsilon_a}\right) W_{lm} I_a \left(\frac{M_2}{M_1}\right)\left(\frac{R_1}{r}\right)^{l+1}, \label{eq:v_a} \\
    H_{\rm int} =& \sum_{l,m} H_{{\rm int}, lm}. \label{eq:H_int}
\end{align}
Note $H_a$ is the energy of a mode in the inertial frame while $H_{a, {\rm ns}}$ energy in the NS frame. We will also collectively denote $H_{\rm mode, ns}=\sum_a^+ H_{a, {\rm ns}}$. 
For future convince, we can view $H_{\rm or}^{\rm (N)}=H_{\rm or}^{\rm (N)}(r, p_r, \phi, P_\phi, H_{\rm mode}, S_{1z, {\rm mode}}, H_{{\rm int}, lm})$ and then treat $H_{\rm mode} = H_{\rm mode}(b_a, d_a)$ and similarly for $S_{1z, {\rm mode}}$. The interaction energy is viewed as $H_{{\rm int}, lm}=H_{{\rm int}, lm}(b_a, d_a, r)$. This allows us to easily note the relation 
\begin{equation}
    \omega \equiv \dot{\phi} = \frac{\partial H_{\rm or, N}}{\partial P_\phi} = -\frac{\partial H_{\rm or, N}}{\partial S_{1z, {\rm mode}}} \simeq  \frac{P_\phi}{\mu r^2}.
\end{equation}
The equation of motion for $b_a$ is 
\begin{align}
    \dot{b}_a &= \frac{\partial H_{\rm or}^{\rm (N)}}{\partial H_{\rm mode}} \frac{\partial H_{\rm mode}}{\partial d_a} + \frac{\partial H_{\rm or}^{\rm (N)}}{\partial S_{1z, {\rm mode}}} \frac{ \partial S_{1z, {\rm mode}}}{\partial d_a} + \frac{\partial H_{\rm or}^{\rm (N)}}{\partial H_{{\rm int}, lm}} \frac{\partial H_{ {\rm int}, lm}}{\partial d_a}, \nonumber \\
    &= - i \left[\omega_a - m_a (\omega - \Omega_1)\right] b_a + i \omega_{a0} v_a.
    \label{eq:db_a_N}
\end{align}
Note from Eqs. (\ref{eq:db_a_N}) and (\ref{eq:dq_a_in}) that an interaction of the form $\Omega S_a$ in the Hamiltonian leads to an effective frequency shift of the mode. In other words, it produces a \emph{frame-dragging} effect (cf. the PN frame-dragging term to be introduced later in Eq.~\ref{eq:H_t_LS}). Note further that the $-P_\phi S_a/(\mu r^2)$ term in our Eq.~(\ref{eq:H_or_N}) is the Newtonian limit of $\mu f_{\rm frame}$ in eq. (5.23) of \cite{Steinhoff:16} (relativistic corrections will be introduced later), and it is the key leading to mode resonance. 
Indeed, if the mode is in equilibrium with the orbit with its phase evolving as $q_a \sim e^{-im_a \phi}$, then $\dot{b}_a$ is small and the equilibrium solution of the mode is 
\begin{equation}
    b_a^{\rm (eq, N)} \simeq \frac{\omega_{a0} v_a}{\omega_a - m_a(\omega - \Omega_1)} = \frac{\omega_{a0} v_a}{\omega_{a0} - m_a \left[\omega - (1 + C_a) \Omega_1\right]},
    \label{eq:b_a_eq_N_nonres}
\end{equation}
where the second equality follows Eq.~(\ref{eq:domega_a_coriolis}). This equation has high accuracy as long as the mode is not near resonance, which happens when the denominator vanishes, or when
\begin{equation}
    \Delta_a \equiv \omega_a - m_a(\omega-\Omega_1) = 0, \text{ or }
    \omega_{\rm res} = \frac{\omega_a}{m_a} + \Omega_1 = \frac{\omega_{a0}}{m_a} + (1 + C_a) \Omega_1.
    \label{eq:res_cond_N}
\end{equation} 
We have used the subscript ``res'' to denote resonance. 

For future convenience, we introduce the effective Love number [for a specific $(l, m)$ harmonic], a multiplicative factor describing the amplification of the NS mass multipoles due to the FF response of a mode,  as 
\begin{align}
    \kappa_{lm} = \frac{1}{2}\left(\frac{b_{a+}}{v_a} + \frac{b_{a-}}{v_a}\right),
    \label{eq:kappa_lm_from_mode}
\end{align}
where $b_{a\pm}$ are amplitudes of f-modes with $(l_a, m_a)=(l, m)$ and $s_a=\pm$. The expression above considers only the f-modes; the general form for $\kappa_{lm}$ including all the NS modes is given in Appx. \ref{appx:q_a_to_multipoles}. 
The total effective Love number used in, e.g., eq. (6) of \cite{Hinderer:16} or eq. (6.9) of \cite{Steinhoff:21}, is given by summing all the $m$ harmonics at a given $l$ as
\begin{equation}
    \kappa_{l, {\rm eff}, r} = - \frac{Q^{\la L \ra}_{\rm mode} \mathcal{E}_{\la L\ra} }{\lambda_l \mathcal{E}^{\la L\ra} \mathcal{E}_{\la L\ra}} = \frac{(2l+1)}{4\pi}\sum W_{lm}^2 \frac{\kappa_{lm}}{\kappa_{l0}}.
    \label{eq:kappa_eff_r}
\end{equation}
We have used $L$ as a shorthand notation for a collection of $l$ individual indices $i_1 ... i_l$, and ``$\la ...\ra$'' to denote a symmetric, trace-free (STF) tensor. We denote the tidal potential as $\mathcal{E}_{\la L\ra} = (2l-1)!! M_2n_{\la L\ra}/r^{l+1}$, where $n^i$ is the Cartesian component of $\vect{n}=\vect{r}/r$ with $\vect{r}$ pointing from the center of $M_1$ towards $M_2$. The NS mass multipole moments induced by the tide are then denoted as $Q^{\la L\ra}_{\rm mode}$. 
The adiabatic tidal deformability $\lambda_l$ follows the definition of eq. (4) in \cite{Flanagan:08} and is given by (see Appx.~\ref{appx:q_a_to_multipoles} for more details)
\begin{equation}
    \lambda_l =\frac{N_l}{l!}R_1^{2l+1} I_a^2 \left(\frac{M_1^2/R_1}{\omega_{a0}\epsilon_a}\right),
    \label{eq:lambda_l}
\end{equation}
where $N_l=4\pi l!/(2l+1)!!$ \cite{Poisson:14}. 
A subscript $r$ is appended in $\kappa_{l, {\rm eff}, r}$ because replacing the adiabatic $\lambda_l$ by $\kappa_{l, {\rm eff}, r} \lambda_l $ preserves the radial tidal interaction \cite{Yu:24a} (see later in Eq. \ref{eq:kappa_eff_h_vs_kappa_eff_r} for a different effective Love number that preserves the GW radiation).

\subsection{Tidal spin and back-reaction torque}
\label{sec:tidal_spin}

\begin{figure}
    \centering
    \includegraphics[width=0.75\linewidth]{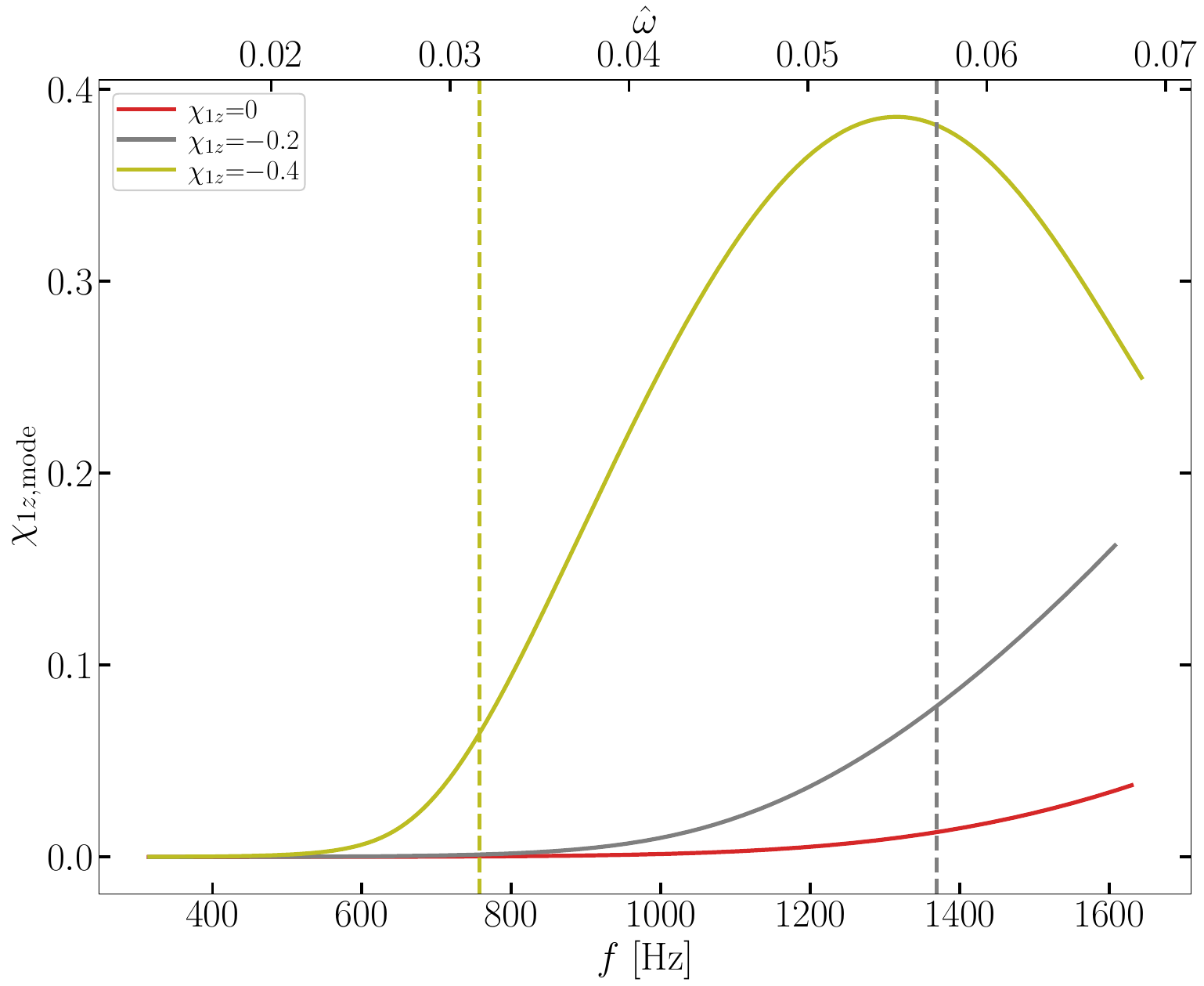}
    \caption{Dimensionless tidal spin for NSs with different background spins. Each curve is terminated when the equal-mass binary contacts. The vertical dashed lines show the resonance frequency of the $l_a=m_a=2$ f-mode for each NS model. When the background spin is zero (red), f-mode resonance does not happen during the inspiral. }
    \label{fig:chi1_t}
\end{figure}

The significance of a coupling with the form $\Omega S_a$ in the mode dynamics has been well recognized in previous studies \cite{Steinhoff:16, Steinhoff:21}. However, the magnitude of such terms was treated as small and the back-reaction on the orbit from it was consequently ignored or incorrectly modeled. 

In the limit where $\omega_{a0} \to \infty$ (i.e., the adiabatic limit), we have $b_a \simeq v_a$ and $H_{a, {\rm ns}}\simeq H_{b, {\rm ns}}$ for two modes with $m_a=-m_b$ (and other quantum numbers the same). Therefore, $S_a + S_b\simeq0$ and the net tidal spin is nearly 0. Consequently, previous analysis \cite{Vines:11, Damour:12} typically ignores the tidal spin. Nonetheless, with FF accounted, we can estimate the net tidal spin in the equilibrium, non-resonance case from Eq.~(\ref{eq:b_a_eq_N_nonres}) as 
\begin{align}
    S_{1z, {\rm mode}}^{\rm (eq)}= \sum_a^+ S_a^{\rm (eq)} \simeq \sum_a^{s_a>0, m_a>0} 4 m_a^2 \epsilon_a \frac{\omega_{a0}^3 [\omega - (1 + C_a)\Omega] v_a^2 }{\left(\omega_{a0}^2 - m_a^2 [\omega - (1 + C_a)\Omega_1]^2\right)^2}, 
    \label{eq:S_1_mode_eq}
\end{align}
where we have paired modes with opposite azimuthal quantum numbers together, so the summation runs over only positive $m_a$. 
Ignoring spin and assume $\omega_{a0} \gg \omega$, we have
\begin{align}
    S_{1z, {\rm mode}}^{\rm (eq)} &\simeq \sum_a^{s_a>0, m_a>0} 4 m_a^2 \epsilon_a \frac{\omega}{\omega_{a0}} v_a^2 \simeq \frac{9 X_2^2}{\omega_{a0}^2} \frac{\lambda_{2} }{M^5} ( M \omega )^5, 
    \label{eq:S_1_mode_low_freq}
\end{align}
where $M=M_1+M_2$ and $X_2=M_2/M$. For future convenience, we will also define $X_1=M_1/M$ and $\eta=X_1X_2$. In the second equality, we have restricted to the $l_a=2$ f-modes and replaced the mode overlap integral $I_a$ to the tidal deformability $\lambda_2$ (Eq. \ref{eq:lambda_l}).
The sharp frequency dependence of $S_{\rm 1,mode} \propto \omega^5$ or $S_{1,{\rm mode}}/p_\phi \propto (M\omega)^{16/3} \equiv x^8$  means it can be an important effect in the late inspiral.\footnote{While the ratio appears like an 8 PN term, we emphasize that the tidal spin appears at the Newtonian order.} It in fact grows faster than $\omega^5$ because of mode resonance (note the denominator in Eq. \ref{eq:S_1_mode_eq} diverges when $\Delta_a=0$). Therefore, whenever the dynamical tide becomes significant, so does the tidal spin, and incorporating the tidal spin would be crucial for improving the faithfulness of analytical BNS/NSBH waveform templates.   

In Fig. \ref{fig:chi1_t}, we present the evolution of the dimensionless tidal spin $\chi_{1z, {\rm mode}} \equiv S_{1z, {\rm mode}}/M_1^2$ obtained from the \emph{full EOB} model up to the point of contact ($r=R_1 + R_2$). We use different colors to indicate different values of the background spin (red, gray, yellow for $\chi_{1z}=0, -0.2, -0.4$, respectively). A negative $\chi_{1z}$ means the \emph{background} spin vector is anti-aligned with the orbital angular momentum. Anti-aligned spins amplify the dynamical tide by lowering the f-mode resonance frequency (shown in vertical dashed lines; GR corrections are included).  For reference,  $\chi_{1z}=0.1$ corresponds to a spin rate of $\Omega_1/2\pi \simeq 195\, {\rm Hz}$ for the SLy EOS adopted in this work, and the maximum spin rate that can be supported is $\Omega_{1, {\rm max}}/2\pi \sim 1.81 \times 10^3\,{\rm Hz}$ \cite{Read:09a}. Near the end of the inspiral, $\chi_{1z, {\rm mode}}$ can reach significant values up to 0.4 (for $\chi_{1z}=-0.4$). Even if the background is non-spinning, the tidal spin can still reach a value of $\chi_{1z, {\rm mode}}=0.04$, which is comparable to the spin prior used when analyzing GW170817 \cite{GW170817, GW170817prop, GW170817eos}. We note that such an evolution of the NS spin is also observed in numerical simulations, e.g., \cite{Kuan:24}. (The amount of evolution in our Fig. \ref{fig:chi1_t} appears to be about twice as large as the one in fig. 1 of \cite{Kuan:24}. This is likely due to the difference between canonical spin used by us and physical spin extracted in \cite{Kuan:24}; see Appx. \ref{appx:can_vs_phy_spin}.)

The growth of the tidal spin is caused by a tidal torque~\cite{Yu:24a}, 
\begin{equation}
    \dot{S}_{1z, {\rm mode}} = \sum_a^+ 2 m_a \omega_{a0}\epsilon_a v_a {\rm Im}[b_a] \equiv - \mu r g_\phi^{(t)},
    \label{eq:dS1z_mode_dt}
\end{equation}
where the equation of motion of the mode, Eq~(\ref{eq:db_a_N}), has been used, and $g_\phi^{(t)}$ describes a tidal back-reaction torque which we will discuss later (see also eq. 55 of \cite{Yu:24a}). Note that whereas the leading-order equilibrium solution of a mode's amplitude given in Eq.~(\ref{eq:b_a_eq_N_nonres}) is purely real, an imaginary piece appears when corrections from finite $\dot{b}_a$ are incorporated. Using the resummation technique introduced in \cite{Yu:24a}, we have
\begin{equation}
    b_a^{\rm (eq)} = \frac{\Delta_a \omega_{a0} v_a }{\Delta_a^2 - i [m_a\dot{\omega} - \Delta_a (l_a+1) \dot{r}/r ]}, 
    \label{eq:b_a_eq_N_w_im}
\end{equation}
which is well-behaving throughout the entire evolution even when the mode reaches resonance. Note that the imaginary piece is caused by the GW-induced orbital decay, which, after the resummation, becomes an effective damping term of the mode that typically dominates over fluid damping by orders of magnitude (see \cite{Lai:94c}). Such an effective damping causes the tidal bulge to lag behind the companion, thereby creating a torque interaction between the deformed body and the orbit as tidal dissipation does in regular stars and planets~\cite{Ogilvie:14}. 

\begin{figure}
    \centering
    \includegraphics[width=0.75\linewidth]{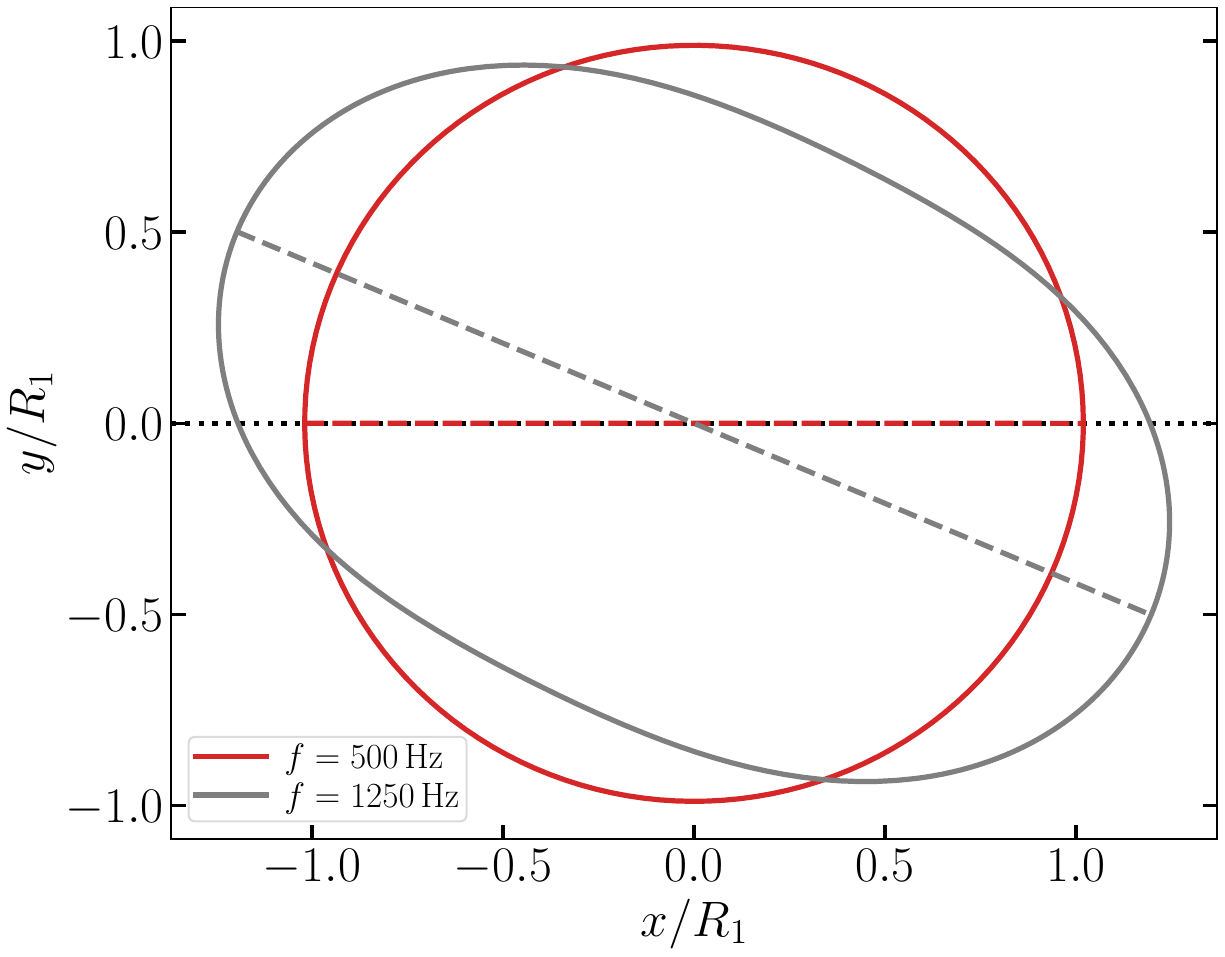}
    \caption{Deformed NS surfaces at two different instants in the orbit frame where the companion is always on the positive x-axis. The red contour is the surface at $f=500\,{\rm Hz}$ and the gray one is at $f=1250\,{\rm Hz}$. For reference, the f-mode resonance is at 1370 Hz for the NS considered here with a dimensionless background spin of $\chi_{1z}=-0.2$. The dashed lines show the major axis. Whereas in the adiabatic limit (approximately true when $f=500\,{\rm Hz}$) the bulge points towards the companion, it lags behind the companion when the tidal response becomes dynamical (as shown in the gray contour). The lag enables a tidal torque.}
    \label{fig:bulge}
\end{figure}

Fig.~\ref{fig:bulge} shows the perturbed surface of $M_1$ reconstructed from our EOB solutions\footnote{The Lagrangian displacement used to generate the plot is still estimated from a Newtonian eigenfunction though we corrected the mode's overlap integral and eigenfrequency by their relativistic values when evolving the dynamics. This affects only the size of the bulge, but not its orientation, which is determined from the phase of $b_a$ and $b_a$ is determined from our EOB model. } in the orbit frame where the companion $M_2$ lies always on the positive $x$ axis. Note the lag increases as the GW frequency increases because the effective damping $(\propto \dot{\omega}/\omega)$ is greater at higher frequencies, and approaches $-\pi/4$ when resonance is reached. As in tides in stars and planets, the torque corresponds to the imaginary part of the Love number \cite{Ogilvie:14}. The authors' previous analysis \cite{Yu:24a} carefully demonstrated that the effective Love number approach commonly adopted by the GW community \cite{Steinhoff:16, Hinderer:16, Steinhoff:21} \emph{does not} capture the torque because the effective Love number is purely real.  

The total AM conservation (in the absence of GW radiation) means the torque driving up the tidal spin will have a back-reaction on the orbit. From the Hamiltonian, Eq.~(\ref{eq:H_or_N}), this is illustrated as 
\begin{equation}
    \dot{P}_\phi = \dot{p}_\phi + \dot{S}_{1z, {\rm mode}} = -\frac{\partial H_{\rm or, N}}{\partial \phi} = 0, \text{ or } \dot{p}_\phi = -\dot{S}_{1z, {\rm mode}}. 
\end{equation}
To further see the back-reaction torque's impact on the orbital evolution, we can consider the circular orbit defined by the minimum of the effective potential, $\partial H_{\rm or}^{\rm (N)}/\partial r=0$, which solves
\begin{equation}
    r = \frac{(P_\phi - S_{1z, {\rm mode}})^2}{\mu^2 M} + \frac{r^3 g_r^{(t)}}{M},
\end{equation}
where $\mu g_r^{(t)} = - \partial H_{\rm int}/\partial r$ is the radial tidal acceleration (see, e.g., eq. 54 from \cite{Yu:24a}). 
Differentiating the above equation and focusing on the torque to the lowest order in the tide, we have
\begin{equation}
    \dot{r}|_{\rm torque} \simeq - \frac{2 P_\phi \dot{S}_{1z}}{\mu^2 M}\simeq \frac{2 g_\phi^{(t)}}{\omega}. 
    \label{eq:drdt_torque}
\end{equation}
The last equality can also be derived from an energy-balancing argument from eq. (62) of \cite{Yu:24a}. 

Such a torque, while missing in the effective Love number prescription, can in principle be captured by the full Hamiltonian described in \cite{Steinhoff:16} (also \cite{Hinderer:16}). However, the authors of \cite{Steinhoff:16} assumed $S_{1z, {\rm mode}}$ to be small and replaced the $-P_\phi S_a /\mu r^2$ term in Eq. (\ref{eq:H_or_N}) by $- M^{1/2} S_a/r^{3/2}$ (see their eq. 6.23), 
\begin{equation}
    H_{\rm or, wrong} = \frac{p_r^2}{2 \mu} + \frac{P_\phi^2}{2 \mu r^2} - \frac{\mu M}{r} + \sum_a^+ \left[H_{a} - \frac{u^{3/2}}{M} S_a + H_{a,{\rm int}} \right],
    \label{eq:H_or_wrong}
\end{equation}
where $u=M/r$.
In other words, they approximated the $P_\phi$ by its pp value $p_\phi$ and then replaced $p_\phi$ by the circular-orbit relation $p_\phi = \mu \sqrt{M r}$. Replacing a canonical variable with another at the Hamiltonian level will not lead to the correct equation of motion, and this can be seen explicitly by again considering the circular orbit found through $\partial H_{\rm or, wrong}/\partial r=0$, leading to 
\begin{align}
    r_{\rm wrong} = \frac{p_\phi^2}{\mu^2 M} - \frac{3}{2}\frac{p_\phi S_{1z, {\rm mode}}}{\mu^2 M} + \frac{r^3 g_r^{(t)}}{M}. 
\end{align}
The evolution driven by the tidal torque from the wrong Hamiltonian is now given by 
\begin{equation}
    \dot{r}|_{\rm torque,\ wrong}=- \frac{3}{2}\frac{p_\phi \dot{S}_{1z, {\rm mode}}}{\mu^2 M} = \frac{3}{4} \dot{r}|_{\rm torque}. 
    \label{eq:drdt_torque_wrong}
\end{equation}
This means even the full EOB model solved by \cite{Steinhoff:16, Hinderer:16} will not correctly capture the tidal torque at the Newtonian order. \cite{Steinhoff:16} did explicitly comment on the error due to replacing $p_\phi$ as a function of $r$ at the Hamiltonian level, and argued it could be justified if $S_{1z, {\rm mode}}$ is sufficiently small. While this could be a reasonable approximation for non-spinning NSs, it loses accuracy in cases where either the NS is rapidly spinning (as demonstrated in Fig. \ref{fig:chi1_t}) or the orbit is eccentric. Either scenario can lead to a strong dynamical tide excitation and a significant value of the tidal spin, and modeling the back-reaction torque correctly would be necessary.

\subsection{PN Hamiltonian}
\label{sec:H_PN}

After describing the key effects of the tidal spin and the need to further improve the EOB model developed in \cite{Steinhoff:16, Steinhoff:21} at the Newtonian level, we now upgrade the dynamics to include PN corrections. In the next section, the PN effects will be resummed into an EOB form which we solve to provide the final waveform. 
Note that throughout the work, we will keep most of the quantities introduced in Secs.~\ref{sec:H_ns_N}-\ref{sec:tidal_spin} with their Newtonian forms and write PN corrections explicitly (either as a multiplicative factor or as new terms appended to the Hamiltonian). For example, $\omega_{a}$ will still mean the eigenfrequency of mode $a$ in an \emph{isolated} NS, and $H_{a, {\rm ns}}$ will still be expressed as $H_{a, {\rm ns}} = \epsilon_a \omega_a q_{a, {\rm in}}^\ast q_{a, {\rm in}}$ as in the Newtonian case. 
The mode amplitudes ($q_a$ or $b_a$) will be the only exception and from this point onward, they will stand for the PN-corrected amplitudes. 
We will work in the inertial frame when applying the relativistic upgrades and apply the generator in Eq.~(\ref{eq:gen_in2or}) to change to the orbit frame in the end. 

Many of the details of incorporating the PN corrections have been worked out in \cite{Steinhoff:16, Steinhoff:21} in terms of the NS mass quadrupole $Q^{ij}_{\rm mode}$. We can therefore make a shortcut and utilize the results from \cite{Steinhoff:16, Steinhoff:21} by relating the mode amplitudes $q_{a}$ used in the Newtonian derivations above to $Q^{ij}_{\rm mode}$. When restricting to the $l=2$ f-modes, the explicit mapping is 
\begin{equation}
    Q^{ij}_{\rm mode} = N_2 M_1 R_1^2 \sum_{m, s} \mathcal{Y}_{2m}^{ij \ast} I_a q_a,  \label{eq:Qij}
\end{equation}
where $\mathcal{Y}_{2m}^{ij}$ is an STF tensor related to the scalar spherical harmonics as \cite{Poisson:14},
\begin{equation}
    Y_{lm} = \mathcal{Y}_{lm}^{\ast \la L \ra} n_{\la L \ra}.
\end{equation}
More general connections including all the eigenmodes and general $l$ will be presented in Appx. \ref{appx:q_a_to_multipoles}. 
Also needed is the momentum conjugate to $Q^{ij}_{\rm mode}$, 
\begin{equation}
    P^{ij} = \frac{1}{2} \frac{M_1}{R_1^3} \sum_{m, s} \left(\frac{\omega_{a0}\epsilon_a}{M_1^2/R_1}\right)  \mathcal{Y}_{2m}^{ij} \frac{i q_a^\ast}{\omega_{a0} I_a}, 
    \label{eq:Pij}
\end{equation}
and the relation between tidal overlap $I_a$ and the tidal deformability $\lambda_{2}$, 
\begin{equation}
    \lambda_2 =\frac{2N_2}{2!}R_1^{5} I_a^2 \left(\frac{M_1^2/R_1}{\omega_{a0}\epsilon_a}\right).
\end{equation}
From $Q^{ij}$ and $P_{ij}$, one can compute the tidal spin as 
\begin{align}
    S_{1z, {\rm mode}} &= \frac{1}{2} \varepsilon_{zij}S_{\rm mode}^{ij} =  2 \varepsilon_{zij} Q_{\rm mode}^{k[i}P^{j]}_{\ k}, \nonumber \\
    &= \sum_{a} m_a \epsilon_a q_a^\ast q_a,
    \label{eq:Sij}
\end{align}
where $\varepsilon$ is the Levi-Civita symbol and ``$[...]$'' around indices represents antisymmetrization. The last equality follows by substituting expressions for $Q^{ij}_{\rm mode}$ and $P_{ij}$ from Eqs.~(\ref{eq:Qij}) and (\ref{eq:Pij}). Thus, the spin tensor used in \cite{Steinhoff:16, Steinhoff:21} is equivalent to the canonical spin introduced in our Eq.~(\ref{eq:S_a_can}). 

Following \cite{Steinhoff:21}, we highlight two PN effects that are most significant for the dynamical tide. The first is a gravitational redshift whose origin can be understood as follows. The total PN action takes a form 
\begin{equation}
    \mathcal{S} \simeq  \mathcal{S}_{\rm pp} + 
    \int  (\sum_a^+ p_{a} \dot{q}_{a} - H_{\rm t, in})d\tau =\mathcal{S}_{\rm pp} + \int z_{\rm pn} ( \sum_a^+ p_{a} \dot{q}_{a} - H_{\rm t, in}) dt,
\end{equation}
where $\mathcal{S}_{\rm pp}$ denotes the PN action for the non-tidal pp part (e.g., \cite{Khalil:20}) and $H_{\rm t, ns}$ is from Eq.~(\ref{eq:H_t_in}). That is, the tidal physics takes the same form as Eq.~(\ref{eq:H_t_in}) when described in terms of $\tau$, the proper time of the worldline along which the dynamical tide propagates. The redshift $z_{\rm PN}$ appears when changing from the proper time to the coordinate time $t$ of an observer at spatial infinity, with
\begin{equation}
    z_{\rm PN}\equiv \frac{d\tau}{dt}.  
\end{equation}
It can be further computed from~\cite{Blanchet:13} 
\begin{align}
    z_{\rm PN} &= \frac{\partial H_{\rm pp}^{\rm (PN)}}{\partial M_1} \simeq 1 - \frac{p_r^2 + p_\phi^2/r^2}{2 M_1^2} - \frac{M_2}{r}, \nonumber \\
    &=1 + \frac{x}{2}(\eta - 3 X_2)\quad \text{(circular orbit)}.
    \label{eq:z_pn}
\end{align}
where in the second line, we have taken the circular orbit limit. We keep only the 1 PN term here, which is sufficient to illustrate the key features introduced by PN. 
Higher PN corrections will be given in the next section directly in terms of the EOB parameter $u=M/r$. The redshift corrects the tidal Hamiltonian $(=-\partial \mathcal{S}/ \partial t)$ as $H_{\rm t, ns}\to z_{\rm PN} H_{\rm t, ns}$, which causes the mode frequency to be redshifted to a lower value $\omega_a \to z_{\rm PN} \omega_a$ as the orbit shrinks. 

The other key modification is relativistic interactions involving the tidal spin $S_{1z, {\rm mode}}$. As argued in \cite{Steinhoff:16, Steinhoff:21} and demonstrated in \cite{Mandal:23}, the tidal spin will enter the PN interaction in exactly the same way as the background, black hole (BH)-like spin $S_{1z}=\chi_{1z} M_1^2$. Using classic results discussing spin interactions in binary black holes (BBHs), we identify spin-orbit and spin-spin interactions involving $S_{1z, {\rm mode}}$ as \cite{Damour:01, Buonanno:06},
\begin{align}
    &H_{t, LS}^{\rm (PN)} = \left(2+\frac{3}{2}\frac{X_2}{X_1}\right)\frac{p_\phi S_{1z,{\rm mode}}}{r^3} + (\text{higher PN terms}), 
    \label{eq:H_t_LS}\\
    &H_{t, SS}^{\rm (PN)} = \frac{\partial H_{{\rm pp,} SS}}{\partial S_1} S_{1z,{\rm mode}} = -\frac{1}{r^3} \left(S_{2z} + \frac{X_2}{X_1}S_{1z}\right) S_{1z, {\rm mode}},
    \label{eq:H_t_SS}
\end{align}
where we have used a ``t'' in the subscripts to indicate the tidal origin of these terms. In particular, the spin-spin term is taken from eq. (2.17) from \cite{Buonanno:06} and we have used $S_{1z} + S_{1z, {\rm mode}}$ as the total spin for $M_1$. The result is then linearized in $S_{1z, {\rm mode}}$.
We have assumed that both spins are aligned. The result is also consistent with eq. (4.4) of \cite{Steinhoff:21}, yet in their eq. (4.8) the contribution from the background spin $S_{1z}$ seemed to be ignored.

Putting things together, the PN-corrected Hamiltonian reads 
\begin{equation}
    H_{\rm in}^{\rm (PN)} = H_{\rm pp}^{\rm (PN)} + z_{\rm pn} H_{\rm mode} + H_{\rm int}^{\rm (PN)} + H_{t, LS}^{\rm (PN)} + H_{t, SS}^{\rm (PN)}.
\end{equation}
Note that the interaction energy is corrected by not only the redshift but also the PN corrections to the tidal potential, so we denote it as $H_{\rm int}^{\rm (PN)}(\neq z_{\rm PN} H_{\rm int}$). The equation of motion for a mode now becomes
\begin{align}
    \dot{q}_a &=\frac{H_{\rm in}^{\rm (PN)}}{\partial p_a} = \{q_a, H_{\rm in}^{\rm (PN)}\},\nonumber  \\
    &=- i\left[z_{\rm PN}\omega_{a} + m_a z_{\rm PN}\Omega + m_a \Omega_{\rm FD}\right]q_a + i \omega_{a0} \left[v_a + dv_a^{\rm (PN)}\right] e^{-i m_a \phi}, 
    \label{eq:q_a_eom_pn}
\end{align}
where
\begin{equation}
    \frac{\Omega_{\rm FD}}{\omega} = \frac{\eta + 3 X_2}{2}x - (\eta \chi_{1z} + X_2^2 \chi_{2z}) x^{3/2}, \quad\text{(circular orbit)}
    \label{eq:Omega_FD}
\end{equation}
and we have lumped all the PN corrections to the driving force into $dv_a^{\rm (PN)}$ which we will give explicit form in the next section directly in terms of the EOB variables. 

The generator in Eq. (\ref{eq:gen_in2or}) still applies if we want to transform the Hamiltonian to the orbit frame. If we just focus on the mode at the moment, the frame transformation can also be done by replacing $q_a=b_a e^{-im_a\phi}$ directly in Eq.~(\ref{eq:q_a_eom_pn}). This allows us to solve for the equilibrium solution as (cf. Eq.~\ref{eq:b_a_eq_N_nonres})
\begin{align}
    b_a^{\rm (eq, PN)}&\simeq \frac{\omega_{a0} \left[v_a + dv_a^{(\rm PN)}\right]}{z_{\rm PN} \omega_a - m_a [\omega - z_{\rm PN}\Omega - \Omega_{\rm FD}] }=\frac{\omega_{a0} \left[v_a + dv_a^{(\rm PN)}\right]}{z_{\rm PN} \omega_{a0} - m_a [\omega - z_{\rm PN} (1 + C_a)\Omega - \Omega_{\rm FD}] }.
    \label{eq:b_a_eq_PN_nonres}
\end{align}
We will collectively denote the denominator as 
\begin{equation}
    \Delta_a^{\rm (PN)}  =z_{\rm PN} \omega_{a0} - m_a [\omega - z_{\rm PN} (1 + C_a)\Omega - \Omega_{\rm FD}]. 
    \label{eq:Delta_a_PN}
\end{equation}
The resonance condition is now given by (cf. Eq.~\ref{eq:res_cond_N})
\begin{align}
    & \Delta_a^{\rm (PN)} = 0 \text{ or }m_a \omega_{\rm res} = z_{\rm PN}(\omega_a + m_a \Omega) + m_a \Omega_{\rm FD}, \nonumber \\
    \text{or } &\omega_{\rm res} \simeq \frac{z_{\rm PN}}{1-\frac{\Omega_{\rm FD}}{\omega_{\rm res}}} \frac{\left[\omega_{a0} + m_a(1+C_a)\Omega_1\right]}{m_a} \simeq \frac{\left[\omega_{a0} + m_a (1+C_a)\Omega\right]}{m_a} \left[1 + \eta x - (\eta \chi_{1z} + X_2^2 \chi_{2z})x^{3/2} \right], 
    \label{eq:res_cond_PN}
\end{align}
which matches eq. (5.12) of \cite{Steinhoff:21} except for that we have an additional $\chi_{1z}$ contribution from the background spin of $M_1$.

Using Eq.~(\ref{eq:b_a_eq_PN_nonres}), we have
\begin{equation}
    \kappa_{lm}^{\rm (eq, PN)} \simeq \frac{z_{\rm PN} \omega_{a0}^2[1 + dv_a^{\rm (PN)}/v_a]}{z_{\rm PN}^2 \omega_{a0}^2 - m_a^2 [\omega - z_{\rm PN}(1+C_a) \Omega_1 - \Omega_{\rm FD} ]^2},
\end{equation}
which is consistent with eq. (5.3) of \cite{Steinhoff:21}.

\begin{figure}
    \centering
    \includegraphics[width=0.75\linewidth]{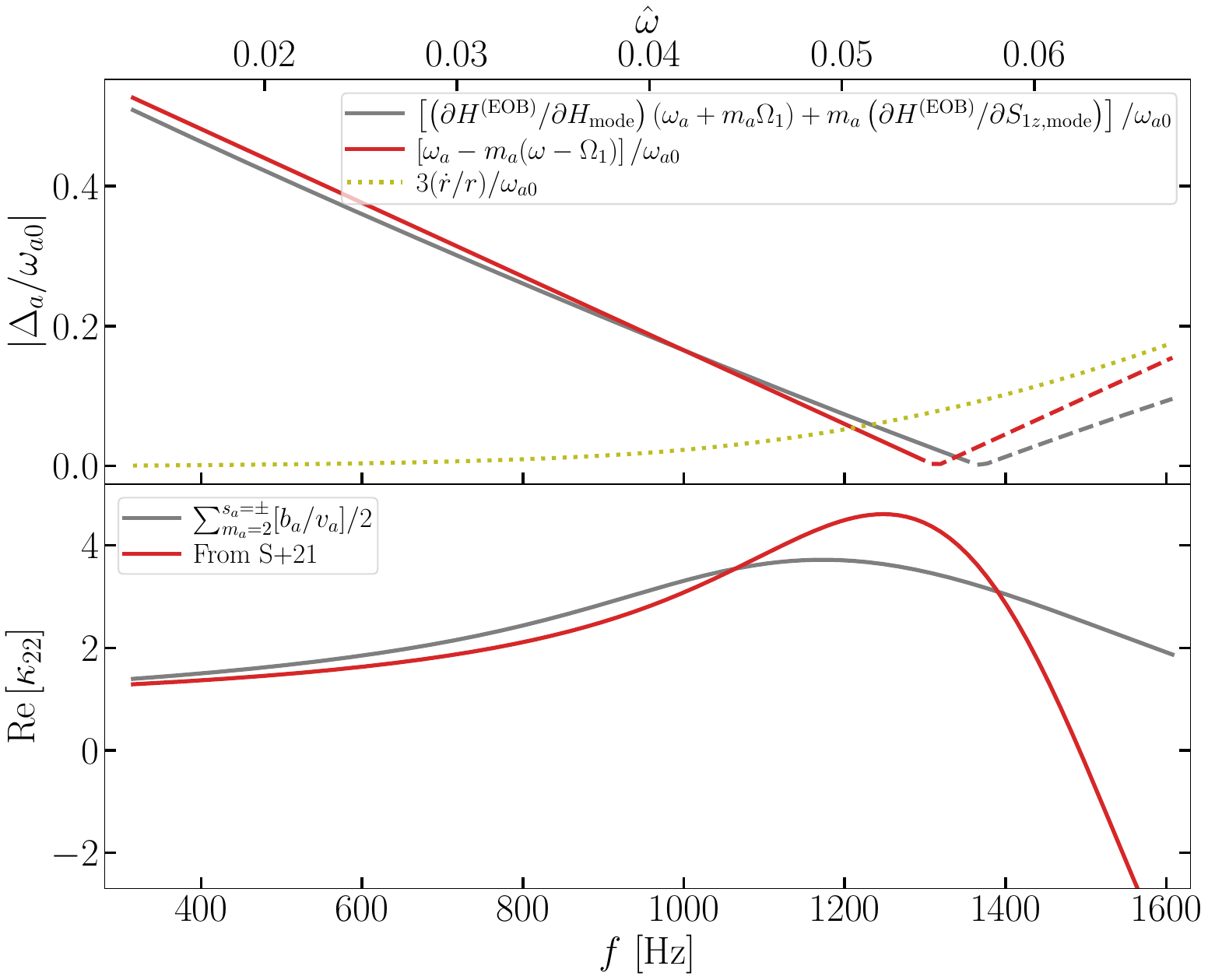}
    \caption{Top: detuning of $m_a=2$ f-mode in a NS with background spin of $\chi_{1z}=-0.2$. The gray line includes the relativistic corrections while the red is obtained under Newtonian physics. The solid (dashed) part is where $\Delta_a$ is positive (negative). Also shown in the yellow-dotted line is the characteristic size of the effective damping induced by the orbit decay. Bottom: FF amplification of the $m_2=2$ NS mass quadrupole. Compared to the Newtonian value used in \cite{Steinhoff:21} (red), the fully relativistic value is initially stronger due to the redshift but weaker near mode resonance as the frame-dragging term becomes dominant. 
    }
    \label{fig:GR_freq_shift}
\end{figure}

Fig.~\ref{fig:GR_freq_shift} shows how the PN corrections modify the detuning of a mode from resonance (top panel) and hence the effective Love number $\kappa_{lm}$ (bottom panel; Eq. \ref{eq:kappa_lm_from_mode}). A background spin of $\chi_{1z}=-0.2$ is assumed. The gray curves are from the full EOB model yet they show good agreement with the leading-order PN corrections. The red curves are from \cite{Steinhoff:21}. Note that while \cite{Steinhoff:21} discussed various PN corrections, the effective Love number they used in the numerical code was actually evaluated at the Newtonian order. In the early inspiral stage, the redshift is the main PN correction, which amplifies the FF response compared to the Newtonian result. Near resonance, the frame-dragging term dominates and shifts the f-mode frequency higher, reducing the response. As demonstrated in \cite{Yu:24a}, the equations in \cite{Steinhoff:21} lose accuracy post resonance. 
In the top panel, we also show in the yellow-dotted line the characteristic size of the effective damping caused by GW decay (see Eq. \ref{eq:b_a_eq_N_w_im}). 

\subsection{EOB Hamiltonian}
\label{sec:eob_hamiltonian}

The PN-expanded Hamiltonian introduced in the previous section can be resummed into an EOB form~\cite{Buonanno:99} which has been shown to be able to capture the strong field dynamics with high accuracy in the case of BBHs. As long as the matter effects remain small, the EOB formulation can also be a good approximation for BNSs. 
In particular, the EOB dynamics can be derived from an effective action resembling a particle moving along geodesic,
\begin{equation}
    \mathcal{S}_{\rm eff} = - \int m_0 d\tau, \text{ with } d\tau = \sqrt{-g_{\mu \nu, {\rm eff}} dx^\mu dx^\nu },
\end{equation}
where $g_{\mu\nu, {\rm eff}}$ is an effective metric along which the effective particle moves. In the test-particle limit, the effective metric reduces to the Schwarzschild or Kerr metric, thereby capturing the strong-field effects in a natural way.
The mass of the particle is $m_0^2= \mu^2 + \mu_{\rm NG}^2$, where $\mu_{\rm NG}$ incorporates non-geodesic effects such as the finite size of the NS as in \cite{Steinhoff:16}. 
The mass shell constraint states
\begin{equation}
    m_0^2 + g_{\rm eff}^{\mu\nu}\frac{\partial \mathcal{S}_{\rm eff}}{\partial x^\mu} \frac{\partial \mathcal{S}_{\rm eff}}{\partial x^\nu} 
    = \mu^2 + \mu_{\rm NG}^2 + g_{\rm eff}^{\mu\nu} p_\mu p_\nu = 0. 
    \label{eq:mass_shell}
\end{equation}
Inverting the equation above leads to the effective, inertial frame Hamiltonian 
\begin{align}
    &H_{\rm in}^{\rm (eff)} = -p_0 = \sqrt{A}\sqrt{\mu^2 + \mu^2_{\rm NG} + \gamma_{\rm eff}^{ij} p_i p_j} + \beta^i p_i, \\
    \text{where } & A=-\frac{1}{g_{\rm eff}^{00}}, \quad 
    \beta^i = \frac{g_{\rm eff}^{0i}}{g_{\rm eff}^{00}}, \quad 
    \gamma_{\rm eff}^{ij}  = g_{\rm eff}^{ij} - \frac{g_{\rm eff}^{0i}g_{\rm eff}^{0j}}{g_{\rm eff }^{ij}}. 
\end{align}
This effective Hamiltonian $H_{\rm in}^{\rm (eff)}\simeq \mu + ...$ can be further mapped to a real Hamiltonian of the binary $H_{\rm in}^{\rm (EOB)}\simeq M + ...$ through \cite{Buonanno:99}
\begin{equation}
    H_{\rm in}^{\rm (EOB)} = M\sqrt{1 + 2 \eta \left[\frac{H_{\rm in}^{\rm (eff)}}{\mu} - 1\right]}. 
    \label{eq:H_eob_from_H_eff}
\end{equation}

The incorporation of the tidal correction into $g_{\rm eff}^{\mu \nu}$ and to $\mu_{\rm NG}$ is not unique especially when the tide can be viewed as a perturbation. 
We follow the same prescription as in \cite{Steinhoff:16} and utilize their results when possible (the key differences will be summarized at the end of this section). 
Specifically, terms that are independent of the metric are incorporated into $\mu_{\rm NG}$. 
Those that are quadratic in $p_0\sim \mu$ in the mass shell constraint (Eq.~\ref{eq:mass_shell}) are incorporated into $A$. This includes the Newtonian tidal interaction $H_{\rm int}$ and part of the 1 PN correction.  
And those quadratic in $p_i$ will be resummed into $\gamma_{\rm eff}^{ij}$. 
Such terms lead to $\mathcal{C}_{ij}$ in \cite{Steinhoff:16}, which constitutes the other part of the PN correction to the interaction. 
Lastly, terms linear in $p_0$ and $p_i$ ($\sim p_\phi$) contribute to $\beta_i$, which includes the spin-orbit interaction $H_{t, LS}$. We further incorporate the spin-spin interaction involving the tidal spin as an additive term to the effective Hamiltonian, $H_{\rm eff}\to H_{\rm eff} + H_{t, SS}$. This is consistent with the treatment in \cite{Buonanno:06}. 
More sophisticated incorporation of the spin-spin interaction such as done in \cite{Khalil:20} is left for future analysis.

Under the above assumption, we write the effective Hamiltonian (in the inertial frame) as 
\begin{align}
    H_{\rm in}^{\rm (eff)} &=H_{\rm in}^{\rm (eff, nts)} + H_{t,LS}^{\rm (eff)} + H_{t,SS}^{\rm (eff)}, \nonumber \\
    &= \mu \sqrt{A} \sqrt{1 + \frac{p_\phi^2}{\mu^2 r^2 } + \frac{A p_r^2}{D_{\rm pp}} + \frac{2 z_E H_{\rm mode} }{\mu} + \frac{2 C_{\rm int}}{\mu}  } + H_{t,LS}^{\rm (eff)} + H_{t,SS}^{\rm (eff)}, \nonumber \\
    &=\mu \sqrt{A} \sqrt{1 + \frac{p_\phi^2}{\mu^2 r^2 } + \frac{A }{A_{\rm pp}^2} {p_r^\star}^2+ \frac{2 z_E H_{\rm mode} }{\mu} + \frac{2 C_{\rm int}}{\mu}  } + H_{t,LS}^{\rm (eff)} + H_{t,SS}^{\rm (eff)},
    \label{eq:H_in_eff}
\end{align}
where $H_{\rm in}^{\rm (eff, nts)}$ is the non-tidal-spin (hence ``nts'' in the superscript) part of the effective Hamiltonian. The potential is
\begin{equation}
    A=A_{\rm pp} + 2 z_I \frac{H_{\rm int}}{\mu}, 
\end{equation}
and $p_r^\star$ is a ``tortoised'' radial momentum~\cite{Damour:07} $p_r^\star = (A_{\rm pp}/\sqrt{D_{\rm pp}}) p_r$ we evolve in the code (though we still use $r$ as the radial coordinate). 
The pp parts of the potentials at the leading orders are
\begin{align}
    A_{\rm pp} &= 1 - 2u + 2 \eta u^3 + ...\\
    D_{\rm pp} &= 1- 6 \eta u^2 + ...
\end{align}
where the first line can be found from, e.g., eq. 5.34 of \cite{Buonanno:99} or eq. (2.2a) of \cite{Barausse:12}, and the second line is given by eq. (5.10) of \cite{Steinhoff:16}. 
For the EOB formulation, we use $u=M/r$ instead of $x=(M\omega)^{2/3}$ to represent the PN order as $r$ is a canonical variable in the EOB dynamics. 
The coefficients $z_I$ and $z_E$ have the form $(1+ u dz)$ with $dz\sim \mathcal{O}(1)$ so that $H_{\rm int}$ and $H_{\rm mode}$ remain their Newtonian forms defined in Eqs. (\ref{eq:H_mode}) and (\ref{eq:H_int}). The $C_{\rm int}$ originates from eq. (3.40) of \cite{Steinhoff:16} and corresponds to a 1 PN correction to the interaction energy that is not absorbed by $A$. 

When expanded, the EOB Hamiltonian needs to match the PN one up to a canonical transformation with generator ${\rm g}$,
\begin{equation}
    H_{\rm in}^{\rm (EOB)} = H_{\rm in}^{\rm (PN)} + \{H_{\rm in}^{\rm (PN)}, {\rm g}\} + \frac{1}{2} \{\{H_{\rm in}^{\rm (PN)}, {\rm g}\}, {\rm g}\} + ...
\end{equation}
which can be used to determine the coefficients in the ansatz in Eq.~(\ref{eq:H_in_eff}). 
A canonical transformation is needed as the EOB Hamiltonian is written with some effective canonical variables $(q_{\rm eff}, p_{\rm eff})$ while the PN one is described with physical ones (e.g., phase space variables in the Arnowitt, Deser, and Misner, or the ADM coordinates); see, e.g., sec. VI of \cite{Buonanno:99} for further discussion. We will, however, omit ``eff'' in the canonical variables because Hamilton's equations are valid in any canonical coordinate system. 

Note that with the pp generator (eq. 5.3 of \cite{Steinhoff:16} or eqs. 6.15 and 6.16 of \cite{Buonanno:99}),
\begin{equation}
    {\rm g}_{\rm pp} = r p_r \left[\frac{\eta}{2}\frac{p^2}{ \mu^2} - \frac{(2+\eta)}{2} \frac{M}{r} \right],
\end{equation}
it is sufficient to change the redshift factor $z_{\rm PN} = \partial H_{\rm pp}^{(\rm PN)}/\partial M_1$ to that of EOB (eq. 6.3 of \cite{Damour:12})
\begin{align}
    z_{\rm EOB} = \frac{\partial H_{\rm pp}^{(\rm EOB)}}{\partial M_1} 
    = 1 - \frac{1}{2}(X_1-1)(X_1-3) u + \frac{3}{8} u^2 (X_1 - 1)(X_1^3 -3X_1^2+3X_1 +3) 
    \label{eq:z_eob}
\end{align}
when the circular orbit is used with $p_r=0$ and $p^2 = p_\phi^2/r^2=\mu^2 (u + 3u^2)$. Such a substitution can be justified through again a canonical transformation with the structure of eq. (5.7) of \cite{Steinhoff:16}. 
The value of $z_{\rm EOB}$ then determines the value of $z_{E}$ by requiring they match when $H_{\rm in}^{\rm (EOB)}$ is expanded, 
\begin{equation}
    z_E = 1 + \frac{3}{2} X_1 u + \frac{27}{8} X_1 u^2, 
\end{equation}
which matches eq. (6.16c) of \cite{Steinhoff:16}. 

The PN corrections to the interaction are less important as they do not affect the mode resonance. For them, we simply follow \cite{Steinhoff:16}. In our notation, we have
\begin{align}
    z_I &= [1 - (2X_2 - \eta)u + \frac{5X_1}{28}(33X_1 - 7)u^2 ],\\
    C_{\rm int} &= \sum_{lm}\sum_a^+ [-C_{lm} u (1+3u)] [-\epsilon_a \omega_{a0} (v_a^\ast b_a + v_a b_a^\ast)]  = \sum_{lm} u dz_{C, lm} H_{{\rm int}, lm}  
\end{align}
where 
\begin{align}
    & u dz_{C,lm} = - C_{lm}u(1+3u), \\
    & \text{with }C_{22}=C_{2,-2} = \eta, \quad C_{20}=(-2 + \eta). 
\end{align}
The results hold only for $l_a=2$ modes. The interaction energies from $l_a\geq 3$ modes are smaller than those of the $l_a=2$ by a factor of $(R_1/r)^{2(l_a-2)}=(R_1/M)^{2(l_a-2)} u^{2(l_a-2)}$, and we include them at the Newtonian accuracy. 

For the tidal spin-orbit interaction, we utilize the results of  \cite{Damour:08} (see their eqs. (3.12), (3.15), and (2.8); higher order terms are available in e.g. \cite{Khalil:20}), 
\begin{align}
    H_{t, LS}^{\rm (eff)} = \left\{\left(2 + \frac{3}{2}\frac{X_1}{X_2}\right) 
    - \left[\frac{5}{8}\eta + \left(\frac{9}{8}+\frac{3}{4}\eta\right)\frac{X_1}{X_2}\right]u\right\}\frac{p_\phi S_{1z, {\rm mode}} }{r^3}
    \label{eq:H_t_LS_eff}
\end{align}
Since we only keep $H_{t, SS}^{\rm (eff)}$ to the leading PN order, we do not include additional generators and ignore the difference between the real and effective coordinates (since they differ at 1 PN). In other words, we use $H_{t, SS}^{\rm (eff)}=H_{t, SS}^{\rm (PN)}$, consistent with \cite{Buonanno:06}. 

With the Hamiltonian constructed in the inertial frame, we can further transform it to the orbit frame to eliminate fast oscillations due to the orbital motion. This can still be achieved with the generator introduced in Eq.~(\ref{eq:gen_in2or}), which transforms $(q_a, p_a)$ to $(b_a, d_a)$ for each mode. For the orbit, the canonical momentum conjugate to $\phi$ becomes $P_\phi = p_\phi+ S_{1z, {\rm mode}}$, and the $p_\phi^2$ term in Eq.~(\ref{eq:H_in_eff}) needs to be replaced by $(P_\phi-S_{1z, {\rm mode}})^2$. This transformation is exact and is what we implement in the numerical code. In other words, we use 
\begin{equation}
    H_{\rm or}^{\rm (eff)} = H_{\rm in}^{\rm (eff)}|_{p_\phi = P_\phi - S_{1z, {\rm mode}}},
    \label{eq:H_or_eff}
\end{equation}
in our code. 
Nonetheless, to assist with the comparison with \cite{Steinhoff:16}, we expand the result as 
\begin{align}
    H_{\rm or}^{\rm (eff)} &\simeq H_{\rm in}^{\rm (eff)} |_{P_\phi} - \frac{\partial H_{\rm in}^{\rm (eff)}}{\partial P_\phi} S_{1z, {\rm mode}}\equiv H_{\rm in}^{\rm (eff)} |_{P_\phi} + H_{\rm fr}^{\rm (eff)},
    \label{eq:H_or_eff_expanded}
\end{align}
where $H_{\rm in}^{\rm (eff)} |_{P_\phi}$ means evaluating Eq.~(\ref{eq:H_in_eff}) with $p_\phi^2$ replaced by $P_\phi^2$, and 
\begin{align}
    H_{\rm fr}^{\rm (eff)} \simeq -\frac{A_{\rm pp}}{H_{\rm pp, in}^{\rm (eff)}}\frac{P_\phi S_{1z, {\rm mode}}}{r^2} \simeq -\left(1-\frac{3}{2}u - \frac{9}{8}u^2\right) \frac{P_\phi S_{1z, {\rm mode}}}{\mu r^2}. 
\end{align}
The subscript ``fr'' indicates this is a term arising from changing frames. 
Note that if we treat $P_\phi\simeq p_\phi$ in the frame dragging terms [$H_{t,LS}^{\rm (eff)} + H_{\rm fr}^{\rm (eff)}$] and further replace $p_\phi$ by the circular orbit relation 
\begin{equation}
\frac{p_\phi}{\mu r}=\sqrt{u}\left[1 + \frac{3}{2}u + \left(\frac{27}{8}-\frac{3}{2}\eta\right)u^2\right], 
\end{equation}
we then recover eq. (6.23) of \cite{Steinhoff:16}. However, replacing the canonical momentum $P_\phi$ in terms of $r$ at the Hamiltonian level is incorrect as discussed in Sec.~\ref{sec:tidal_spin} [see Eqs.~(\ref{eq:drdt_torque}) and (\ref{eq:drdt_torque_wrong})]. This is the main difference between this work and \cite{Steinhoff:16}. 

Similarly, the EOB Hamiltonian can be written as 
\begin{equation}
    H^{\rm (EOB)}  \simeq H^{\rm (EOB)}_{\rm in}|_{P_\phi} - \frac{\partial H_{\rm in}^{\rm (EOB)}}{\partial P_\phi} S_{1z, {\rm mode}} = H^{\rm (EOB)}_{\rm in}|_{P_\phi} + H_{\rm fr}^{\rm (EOB)}.
\end{equation}
Here we drop the ``or'' subscript in $H^{\rm (EOB)}$ for notational simplicity. This orbit frame Hamiltonian is the final one we will evolve. Note $\omega = \dot{\phi} = \partial H_{\rm in}^{\rm (EOB)} / \partial P_\phi$, so in the equation of motion for the mode, the $H_{\rm fr}^{\rm (EOB)}$ term creates a frequency shift of the mode of $-m_a \omega$, as expected.

To summarize, the effective Hamiltonian we use is given in Eq.~(\ref{eq:H_or_eff}), which has the same form as Eq.~(\ref{eq:H_in_eff}) but with $p_\phi$ replaced by $(P_\phi - S_{1z, {\rm mode}})$ [or $(P_\phi - S_{1z, {\rm mode}} - S_{2z, {\rm mode}})$ if $M_2$ is also an NS]. The tidal spin-orbit and spin-spin interactions are respectively given by Eq.~(\ref{eq:H_t_LS_eff}) and Eq.~(\ref{eq:H_t_SS}). The effective Hamiltonian is then mapped to the real EOB Hamiltonian according to Eq.~(\ref{eq:H_eob_from_H_eff}) with ``in'' replaced by ``or'' [and ``or'' is further dropped in $H^{\rm (EOB)}$].

The key differences between our model and the full Hamiltonian in \cite{Steinhoff:16} (in their sec. VI B before the effective Love number approximation is adopted) are the following. 
\begin{enumerate}
\item We evolve each mode's amplitude instead of $Q^{ij}_{\rm mode}$ and $P_{ij}$. The two approaches are equivalent when only the f-modes are considered, yet our model is more flexible in incorporating other modes (e.g., gravity and pressure modes). For a spinning background, the configuration-space formulation in \cite{Steinhoff:21} leads to coupled equations of motion for different modes. This is avoided in the phase-space expansion we adopt following \cite{Schenk:02}. More importantly, this work paves the way for including nonlinear hydrodynamic interactions which have been shown to be important in the Newtonian order \cite{Yu:23a}. 

\item We use $P_\phi=p_\phi + S_{1z, {\rm mode}} + S_{2z, {\rm mode}}$ as the canonical momentum conjugate to $\phi$ in the orbit frame instead of the pp one $p_\phi$. 
For the frame dragging terms [$H_{\rm fr}^{\rm (eff)}$ and $H_{t, LS}^{\rm (eff)}$], we \emph{do not} replace $P_\phi$ by $\mu r \sqrt{u}[1+...]$ in order to capture the correct back-reaction torque (as discussed in Sec.~\ref{sec:tidal_spin}). In fact, $H_{\rm fr}^{\rm (eff)}$ does not explicitly show up in our code as we use the exact Eq.~(\ref{eq:H_or_eff}) instead of the expanded Eq.~(\ref{eq:H_or_eff_expanded}). 

\item We included the leading-order tidal spin-background spin interaction (Eq.~\ref{eq:H_t_SS}). Compared to \cite{Steinhoff:21}, we additionally have $S_{1z, {\rm mode}}$ interacting with $M_1$'s own background spin $\chi_{1z}$. 
\end{enumerate}

\subsubsection{Dynamics defined by the EOB Hamiltonian}

In the numerical code, it is convenient to evolve the system in dimensionless units. This is achieved by scaling length and time by $1/M$, frequency by $M$, energy and linear momentum by $1/\mu$, and angular momentum by $1/(\mu M)$. In particular, the pp part of the dynamics is scaled as 
\begin{equation}
    \left(r, p_r, \phi, p_\phi, t, H^{\rm (EOB)}, H_{\rm or}^{(\rm eff)}\right) \to \left(\frac{r}{M}, \frac{p_r}{\mu}, \phi, \frac{p_\phi}{\mu M}, \frac{t}{M}, \frac{H^{\rm (EOB)}}{\mu}, \frac{H_{\rm or}^{(\rm eff)}}{\mu}\right). 
\end{equation}
For the tidal part, we scale 
\begin{equation}
    \left(\epsilon_a, \omega_{a0}, \Omega_{1}, H_{\rm mode}, S_{1z, {\rm mode}}, H_{{\rm int}, lm}\right) \to 
    \left(\frac{\epsilon_a}{\mu M}, M\omega_{a0}, M\Omega_1, \frac{H_{\rm mode}}{\mu}, \frac{S_{1z, {\rm mode}}}{\mu M}, \frac{H_{{\rm int}, lm}}{\mu}\right). 
\end{equation}
This way, $b_a$ and $v_a$ retain their original meanings (as they are already dimensionless). We will denote the scaled, dimensionless quantities by a hat symbol subsequently.

As in Sec.~\ref{sec:frames}, we treat
\begin{equation}
    \hat{H}^{\rm (EOB)} = \hat{H}^{\rm (EOB)} [\hat{r}, \hat{p}_r^\star, \phi, \hat{P}_\phi, \hat{H}_{\rm mode}(b_a, \hat{d}_a), \hat{S}_{1z, {\rm mode}}(b_a, \hat{d}_a), \hat{H}_{{\rm int}, lm}(\hat{r}, b_a, \hat{d}_a)], 
    \label{eq:H_eob_vs_vars}
\end{equation}
This allows us to write the mode's evolution analytically via the chain rule as 
\begin{align}
    \frac{d b_a}{d\hat{t}} = \frac{\partial \hat{H}^{(\rm EOB)}}{\partial \hat{H}_{\rm or}^{\rm (eff)}}\left[\frac{\partial \hat{H}_{\rm or}^{\rm (eff)}}{\partial \hat{H}_{\rm mode}}\frac{\partial \hat{H}_{\rm mode}}{\partial \hat{d}_a} 
    + \frac{\partial \hat{H}_{\rm or}^{\rm (eff)}}{\partial \hat{S}_{1z,{\rm mode}}}\frac{\partial \hat{S}_{1z,{\rm mode}}}{\partial \hat{d}_a}
    + \frac{\partial \hat{H}_{\rm or}^{\rm (eff)}}{\partial \hat{H}_{{\rm int}, lm}}\frac{\partial \hat{H}_{{\rm int}, lm}}{\partial \hat{d}_a}\right],
    \label{eq:db_a_EOB}
\end{align}
where
\begin{align}
    &\frac{\partial \hat{H}^{(\rm EOB)}}{\partial \hat{H}_{\rm or}^{\rm (eff)}}=\frac{1}{\eta \hat{H}^{\rm (EOB)}}, \\
    &\frac{\partial \hat{H}_{\rm or}^{\rm (eff)}}{\partial \hat{H}_{\rm mode}}=\frac{\partial \hat{H}_{\rm or}^{\rm (eff, nts)}}{\partial \hat{H}_{\rm mode}} = \frac{z_E A}{\hat{H}_{\rm or}^{(\rm eff,\ nts)}}, \\
    &\frac{\partial \hat{H}_{\rm or}^{\rm (eff, nts)}}{\partial \hat{S}_{1z, {\rm mode}}} = - \frac{\hat{p}_\phi}{\hat{r}^2}\frac{A}{\hat{H}_{\rm or}^{(\rm eff,\ nts)}}, \\
    &\frac{\partial \hat{H}_{\rm or}^{\rm (eff)}}{\partial \hat{H}_{{\rm int}, lm}}=\frac{\partial \hat{H}_{\rm or}^{\rm (eff, nts)}}{\partial \hat{H}_{{\rm int}, lm}} = z_I \frac{\hat{H}_{\rm or}^{(\rm eff,\ nts)}}{A} + \left(u dz_{C, lm} + z_I \frac{ {\hat{p}_r^\star}{}^2}{A_{\rm pp}^2}\right) \frac{A}{\hat{H}_{\rm or}^{(\rm eff,\ nts)}},
\end{align}
and 
\begin{align}
    &\frac{\partial \hat{H}_{\rm mode}}{\partial \hat{d}_a} = -i (\hat{\omega}_a + m_a \hat{\Omega}) b_a,\quad 
    \frac{\partial \hat{S}_{1z,{\rm mode}}}{\partial \hat{d}_a} = - i m_ab_a, \quad 
    \frac{\partial \hat{H}_{{\rm int}, lm}}{\partial \hat{d}_a} = i \omega_{a0} v_a.
\end{align}

The description is completed by coupling the mode evolution with the orbital part (e.g. \cite{Pan:11}), 
\begin{align}
    &\frac{d\hat{r}}{d\hat{t}} = \frac{A_{\rm pp}}{\sqrt{D_{\rm pp}}} \frac{\partial \hat{H}^{\rm (EOB)}}{\partial \hat{p}_r^\star}, \label{eq:dr_dt}\\
    &\frac{d \hat{p}_r^\star}{d\hat{t}} = - \frac{A_{\rm pp}}{\sqrt{D_{\rm pp}}}\left[\frac{\partial \hat{H}^{\rm (EOB)}}{\partial \hat{r}} 
    -\sum_{lm}(l+1) \frac{\partial \hat{H}^{(\rm EOB)}}{\partial \hat{H}_{\rm or}^{\rm (eff)}} \frac{\partial \hat{H}_{\rm or}^{\rm (eff)}}{\partial \hat{H}_{{\rm int}, lm}} \frac{\hat{H}_{{\rm int}, lm}}{\hat{r}}\right] + \hat{F}_\phi \frac{\hat{p}_r^\star}{\hat{P}_\phi}, \\
    &\frac{d\phi}{d\hat{t}} = \frac{\partial \hat{H}^{\rm (EOB)}}{\partial \hat{P}_{\phi}}= \hat{\omega} \equiv v_\omega^3, \\
    &\frac{d \hat{P}_\phi}{d\hat{t}} = \hat{F}_\phi, \label{eq:dP_phi_dt}
\end{align}
where $\hat{F}_\phi$ describes the dissipative GW radiation which we will discuss later in Sec. \ref{sec:GW_rad}. 

\subsubsection{Equilibrium solutions and initial conditions}

The equilibrium solution of the mode, prior to resonance, can be obtained by setting $d b_a /d \hat{t}=0$, which leads to 
\begin{align}
    &b_a^{\rm (eq, EOB,0)} = \frac{\left[\partial H^{\rm (EOB)}/\partial H_{\rm int}\right]\omega_{a0} v_a}{\Delta_a^{\rm (EOB)}}, \label{eq:b_a_eq_EOB_0}\\
    \text{where } & \Delta_a^{(\rm {EOB})} =\left[\frac{\partial H^{(\rm EOB)}}{\partial H_{\rm mode}}\right](\omega_a + m_a \Omega_1) + m_a \left[\frac{\partial H^{\rm (EOB)}}{\partial S_{1z, {\rm mode}}}\right].
    \label{eq:Delta_a_EOB}
\end{align}
The detuning is to be compared with Eqs.~(\ref{eq:res_cond_N}) and (\ref{eq:Delta_a_PN}). 
We further append a superscript $0$ in the equilibrium mode solution above, because it gives a real $b_a$ and does not capture the lag of the bulge depicted in Fig.~\ref{fig:bulge} due to the imaginary part of $b_a$. To capture this lag, we can substitute Eq. (\ref{eq:b_a_eq_EOB_0}) back to Eq. (\ref{eq:db_a_EOB}) for $db_a/d\hat{t}$, leading to the next-order correction $b_a^{\rm (eq, EOB,1)}= i  \dot{b}_a^{\rm (eq, EOB, 0)} / \Delta_a^{\rm (EOB)}$. 
We then follow \cite{Yu:24a} and perform a resummation to obtain the final equilibrium solution 
\begin{align}
    b_a^{(\rm eq, EOB)} &= b_a^{(\rm eq, EOB, 0)} + b_a^{(\rm eq, EOB, 1)}+... \simeq \frac{b_a^{(\rm eq, EOB, 0)}}{1 - b_a^{(\rm eq, EOB, 1)}/b_a^{(\rm eq, EOB, 0)}},  & \nonumber \\
    &\simeq \frac{\Delta_{a}^{(\rm EOB)}[\partial H^{(\rm EOB)}/\partial H_{{\rm int}}]\omega_{a0} v_a}{\left[\Delta_a^{\rm (EOB)}\right]^2 + i \left[(3/2)m_a\omega + (l_a+1)\Delta_a^{(\rm EOB)} \right](\dot{r}/r)},
    \label{eq:b_a_eq_EOB_w_im}
\end{align}
where we have used the Newtonian solution from Eq.~(\ref{eq:b_a_eq_N_nonres}) to evaluate $\dot{b}_a^{\rm (eq)}$ and approximated $\dot{\omega}/\omega \simeq -(3/2)\dot{r}/r$.

Using the equilibrium solutions, the initial condition of our system is determined iteratively. At a given initial separation $r$, we first find the total angular momentum by numerically searching $P_\phi$ such that
\begin{equation}
    \frac{\partial H^{\rm (EOB)}}{\partial r} - \sum_{lm} (l+1)\frac{\partial H^{\rm (EOB)}}{\partial H_{{\rm int}, lm}}\frac{H_{{\rm int}, lm}}{r}=0, 
\end{equation}
which corresponds to finding the circular orbit solution. Note we have used Eq.~(\ref{eq:H_eob_vs_vars}) which leads to the $\partial H^{\rm (EOB)}/\partial H_{{\rm int}, lm}$ terms. Newtonian mode amplitudes from Eq.~(\ref{eq:b_a_eq_N_w_im}) together with $p_r=0$ are used in the root searching above for evaluating $H^{\rm (EOB)}$.  
We then approximate $\dot{r}\simeq -(64/5)\eta u^3$ and write $\hat{p}_r^\star$ in terms of $\dot{r}$ using Eq. (\ref{eq:dr_dt}), 
\begin{equation}
    \hat{p}_r^\star \simeq \frac{\sqrt{D_{\rm pp}}}{A_{\rm pp}}\left(\frac{A_{\rm pp}}{A}\right)^2 \eta \hat{H}^{\rm (EOB)} \hat{H}^{\rm (eff, nts)} \frac{d \hat{r}}{d\hat{t}}.
\end{equation}
With $(r, p_r^\star, P_\phi, b_a)$ determined so far, we evaluate $\Delta_a^{(\rm EOB)}$ from Eq. (\ref{eq:Delta_a_EOB}) together with
$\omega = \partial H^{\rm (EOB)}/\partial P_\phi$ 
so that we can use Eq.~(\ref{eq:b_a_eq_EOB_w_im}) to update the mode amplitudes,  completing the initial conditions for the system. 

\begin{figure}
    \centering
    \includegraphics[width=0.75\linewidth]{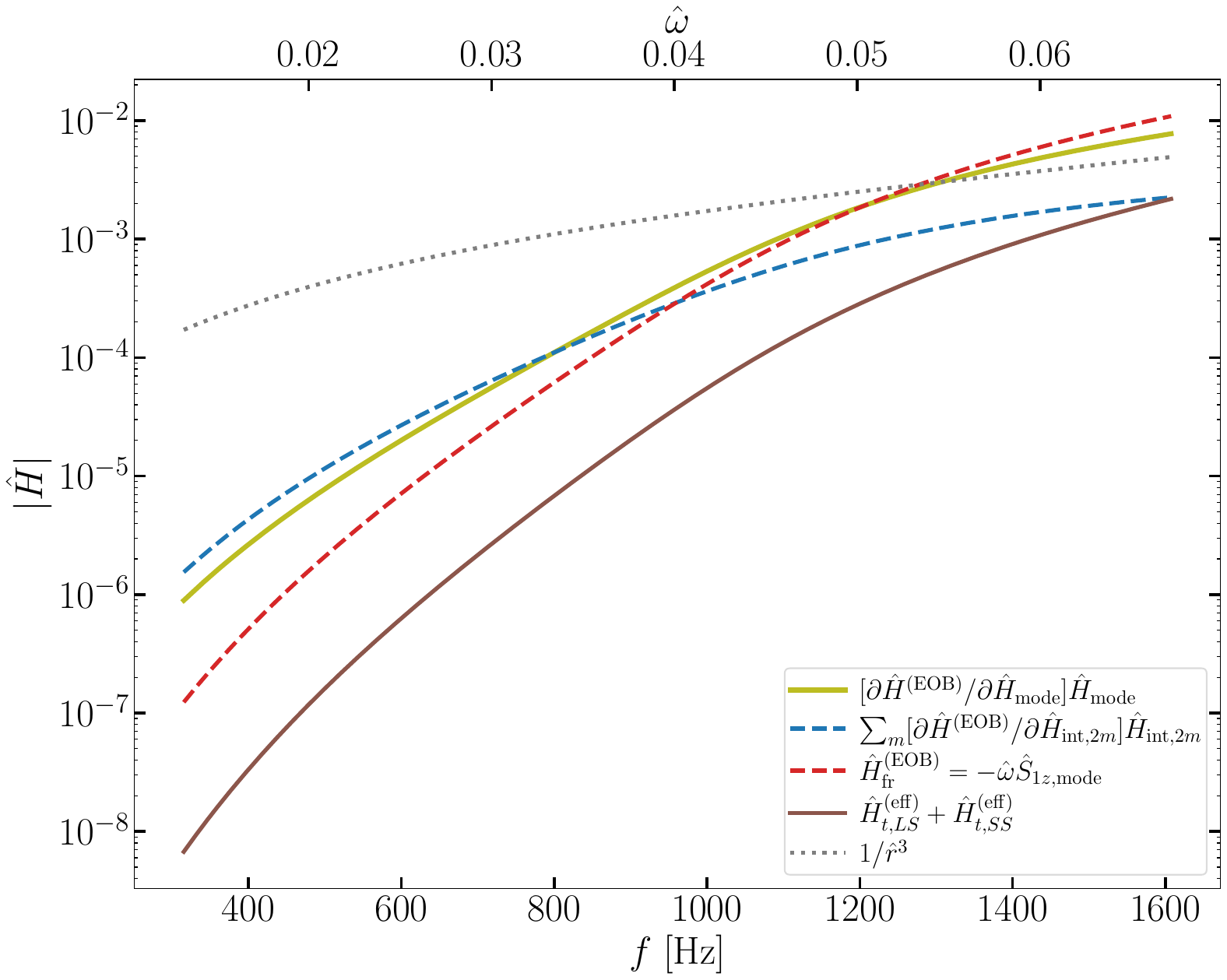}
    \caption{Comparisons of different terms in the expanded EOB Hamiltonian. All terms are normalized by the reduced mass $\mu$. Also shown in the dotted line is the characteristic size of a 3 PN term in the pp (i.e., BBH) Hamiltonian. The tidal spin-related terms (red and brown) can have magnitudes similar to or even greater than the 3 PN pp term. 
    }
    \label{fig:H_comp}
\end{figure}

To conclude our discussions on the conservative dynamics, we linearize the orbit-frame EOB Hamiltonian with respect to the $\lambda$ and then expand it to 1 PN and show its consistency with earlier studies. 
\begin{align}
    H_{t, {\rm EOB}}=&H_{\rm EOB}-H_{\rm pp, EOB} \nonumber \\
    \simeq & \left[1+\frac{u}{2}(-3 + \eta + 2dz_E)  \right]H_{\rm mode} + \sum_{lm}\left[1 + \frac{u}{2}(3+\eta + 2dz_I + 2dz_{C, lm} ) \right] H_{{\rm int}, lm} \nonumber \\
    &+ H_{\rm fr}^{(\rm EOB)} + H_{t, LS}^{\rm (eff)} + H_{t, SS}^{\rm (eff)}, \nonumber \\
    =&\left[1-\frac{u}{2}(X_1-3)(X_1-1)\right]H_{\rm mode} + \left[1 + \frac{u}{2} (7-4X_2 + \eta) \right]H_{\rm int, 20} + 2\left[1+\frac{u}{2}(3 - 4X_2 + \eta)\right] H_{\rm int, 2,|2|} \nonumber \\
    & + H_{\rm fr}^{(\rm EOB)} +  H_{t, LS}^{\rm (eff)} + H_{t, SS}^{\rm (eff)}.
    \label{eq:H_t_EOB_expanded}
\end{align}
We check the coefficient before $H_{\rm mode}$ matches eq. 6.3 of \cite{Damour:12}. The coefficients of $H_{{\rm int}, lm}$ (note the difference between $m=|2|$ and $m=0$) can be checked by getting the equation of motion for $b_a$,  
\begin{align}
    \dot{b}_a =\{b_a, H^{\rm (EOB)}\}\simeq - i \left[z_{\rm EOB}(\omega_a + m_a\Omega) - m_a z_{\rm fr} \frac{u^{3/2}}{M} + m_a \Omega_{\rm FD} \right]b_a + i \omega_{a0} \left[v_a + dv_a^{\rm (EOB)}\right],
    \label{eq:b_a_eom_eob_expanded}
\end{align}
where
\begin{align}
    dv_a^{\rm (EOB)} &=  \frac{u}{2}(7-4X_2 + \eta) v_a,  &\text{for $l_a=2, m_a=0$}, \\
    dv_a^{\rm (EOB)} &= \frac{u}{2}(3-4X_2 + \eta) v_a,  &\text{for $l_a=|m_a|=2$}. 
\end{align}
and $z_{\rm EOB}$ is given by Eq.~(\ref{eq:z_eob}), and $\Omega_{\rm FD}$ by Eq. (\ref{eq:Omega_FD}). 
Further, 
\begin{align}
    z_{\rm fr}\frac{u^{3/2}}{M} =\frac{\partial H_{\rm in}^{\rm (EOB)}}{\partial P_\phi} = \omega \simeq \frac{u^{3/2}}{M} \left(1 + \frac{\eta u}{2}\right). 
\end{align}
The equilibrium configuration is computed by setting $\dot{b}_a=0$, which takes the same form as Eq.~(\ref{eq:b_a_eq_PN_nonres}) prior to resonance. When taking $\omega_{a}\to \infty$, we have the adiabatic limit
\begin{equation}
    b_{a}^{\rm (ad)}=\lim_{\omega_a\to \infty} b_a = \frac{1}{1+u dz_{\rm EOB}} [v_a+dv_a^{\rm (EOB)}],
    \label{eq:b_a_PN_ad}
\end{equation}
which matches eqs. (B2c) and (B3) of \cite{Steinhoff:16} for $m_a=0$ and $m_a=|2|$, respectively.  

In Fig. \ref{fig:H_comp}, we compare various terms in Eq. (\ref{eq:H_t_EOB_expanded}) for an NS with $\chi_{1z}=-0.2$ inspiraling in an equal mass binary. Solid (dashed) curves are for positive (negative) terms. 
We highlight that terms involving the tidal spin (red and brown) both can reach significant magnitudes, emphasizing their significance in the dynamics. The $H_{\rm fr}^{(\rm EOB)}$ term which is closely related to the tidal torque is in fact the dominant tidal term in the Hamiltonian. Its magnitude exceeds a typical 3 PN term in the pp Hamiltonian shown in the dotted line.

\section{Dissipative GW radiation}
\label{sec:GW_rad}

Besides the conservative dynamics discussed above in terms of the Hamiltonian, dissipative effects due to GW radiation are another critical component for a binary's evolution. 
In previous studies, tidal effects in the radiation were included either under the adiabatic limit \cite{Vines:11, Damour:12} or by replacing the adiabatic Love number with the effective one \cite{Steinhoff:16, Hinderer:16, Steinhoff:21}. There are at least two issues when mode resonance is included. 

First, note the energy loss due to GW radiation at the leading order can be written as 
\begin{equation}
    \dot{E} = - \frac{1}{5}\la \dddot{Q}^{ij} \dddot{Q}_{ij} \ra,
    \label{eq:dEdt_vs_quad}
\end{equation}
where the system mass quadrupole is the sum of the pp orbital part and a part induced by the tidal perturbation, $Q^{ij}=Q^{ij}_{\rm orb} + \Delta Q^{ij} $. The tide perturbs the system quadrupole via two channels. One is directly inducing a quadrupole, $Q_{\rm mode}^{ij}$, inside the NS from the tidally driven modes, and the other is through back-reacting on the orbit, which modifies the equilibrium separation at a given frequency~\cite{Yu:23a}. We have
\begin{align}
    \Delta \dddot{Q}^{ij} &= \dddot{Q}^{ij}_{\rm mode} + \dddot{Q}^{ij}_{\rm br}\simeq \dddot{Q}^{ij}_{\rm mode} + 2\left(\frac{\Delta r}{r}\right) \dddot{Q}_{\rm orb}^{ij}. 
\end{align}
In the adiabatic limit, both terms are proportional to $\lambda$ and of the same magnitude. However, when mode resonance is included, the effective Love number replacement (which includes the $m=0$ tide) only correctly preserves the second, back-reaction channel (the $\propto \Delta r/r$ term). The $m=0$ tide does not contribute to the time derivatives of the NS quadrupole and therefore the effective Love number does not give the correct description of the $\dddot{Q}^{ij}_{\rm mode}$ term under mode resonance (see also, \cite{Yu:24a}).\footnote{
In \texttt{LAL} \cite{LAL:18} (at least the version used by \cite{Steinhoff:21}; see their footnote 3), there seems to be a function \texttt{XLALSimIMRTEOBk2effMode} under \texttt{LALSimIMRSpinEOBFactorizedWaveform.c} that attempts to compute a different effect Love number, $\kappa_{{\rm eff}, h}$, for the radiation in terms of the radial Love number [Eq. (\ref{eq:kappa_eff_r})]. However, their implementation appears to be incorrect
\begin{equation*}
    \kappa_{{\rm eff}, h}^{\rm (LAL)} \simeq \frac{(-1 +  \kappa_{{\rm eff},r})(\omega_f/\omega)^2 + 6 X_2 \kappa_{{\rm eff},r}}{3 (1+2X_2)},
\end{equation*}
which does not match our Eq. (\ref{eq:kappa_eff_h_vs_kappa_eff_r}) even at the Newtonian order. 
\label{ft:kappa_h_in_lal}
}

The second issue is that the induced NS mass quadrupole contains two harmonics instead of one in the presence of mode resonance \cite{Yu:24a}. This is a direct consequence that a general forced oscillator's solution is the sum of a specific one (the equilibrium tide varying with the tidal drive as $e^{-im \phi}$) and a homogeneous one (the dynamical tide excited near resonance oscillating at the oscillator's natural frequency). From Eq. (\ref{eq:dEdt_vs_quad}), it is clear that only the equilibrium tide can interact with the orbital quadrupole and have a non-vanishing time average. The dynamical component is incoherent with the orbit (even at resonance when the frequencies match; \cite{Yu:24a}) and does not produce $\mathcal{O}(\lambda)$ correction to the energy loss. However,
previous work did not separate the equilibrium and dynamical components, leading to inaccurate tidal contributions near resonances. 

To overcome the above two issues and self-consistently include mode resonance into the GW radiation, we follow \cite{Vines:13} who derived the global-frame system multipole moments ($Q_{\rm sys}^L$ for mass and $S_{\rm sys}^L$ for current) in terms of general body-frame moments (in particular, $Q^{ij}_{\rm mode}$ and $S^{ij}_{\rm mode}$; Eq. \ref{eq:Qij} and \ref{eq:Sij}) to 1 PN order. Higher PN order results are recently derived in \cite{Mandal:24}, yet we find the 1 PN accuracy is sufficient for the current work.   
The global-frame multipoles are then used to derive the GW modes based on \cite{Blanchet:14}
(an overall minus sign is introduced so that the expression matches the sign convention in \cite{Damour:09})
\begin{equation}
    h_{lm} \simeq  \frac{1}{D_L} \frac{8 \pi}{(2l+1)!!} \mathcal{Y}^{\la L \ra}_{lm} \left[\sqrt{\frac{(l+1)(l+2)}{l(l-1)}}  Q_{{\rm sys}, \la L\ra}^{(l)} 
    +  2i \sqrt{\frac{l(l+2)}{(l+1)(l-1)}} S_{{\rm sys,} \la L\ra}^{(l)}\right] + \text{(GW tails)},
    \label{eq:h_lm_from_multipoles}
\end{equation}
where $D_L$ is the luminosity distance of the source and ``$(l)$'' means taking the time derivative $l$ times. It is also convenient to convert the multipole tensor to a particular $(l, m)$ mode as $Q_{{\rm sys}, lm} = \mathcal{Y}_{lm}^{\la L\ra}  Q_{{\rm sys}, \la L\ra}$ and similarly for $S_{{\rm sys}, lm}$. 
When a Newtonian (indicated by a superscript N), pp orbit is assumed, we have (e.g., \cite{Damour:09, Pan:11})
\begin{align}
    h_{lm}^{\rm (N,\zeta)}=c_{(l+\zeta)} h^{\rm '(N, \zeta)}_{lm} = \frac{\eta M}{D_L} c_{(l+\zeta)} n_{lm}^{(\zeta)} v_\phi^{l+\zeta} Y_{l-\zeta, -m}\left(\frac{\pi}{2}, \phi\right),
    \label{eq:h_lm_pp_N}
\end{align}
where $\zeta = {\rm mod} (l+m,\ 2)$ and 
\begin{align}
    &c_{(l+\zeta)} = X_2 ^{l+\zeta-1} + (-1)^{l+\zeta} X_1^{l+\zeta-1},\\
    &n_{lm}^{(0)} = (im)^l \frac{8\pi}{(2l+1)!!} \sqrt{\frac{(l+1)(l+2)}{l(l-1)}},\\
    &n_{lm}^{(1)} = -(im)^l \frac{16 \pi i}{(2l+1)!!} \sqrt{\frac{(2l+1)(l+2)(l^2-m^2)}{(2l-1)(l+1)l(l-1)}},  \\
    &v_\phi \equiv \hat{\Omega} \left[\frac{\partial \hat{H}^{\rm (EOB)}}{\partial \hat{P_\phi}}\right]_{\hat{p}_r^\star=0}^{-2/3}. 
\end{align}
Note that $v_\phi$, as introduced in \cite{Damour:06}, is different from $\hat{\Omega}^{1/3}$ as the quantity inside the bracket is evaluated by setting $p^\star_r=0$ in the Hamiltonian, though the difference between the two is typically small. 
We will use $h^{\rm '(N, \zeta)}_{lm}$ to normalize the tidal contributions later. 

To capture FF corrections to the f-mode response, we decompose 
\begin{align}
    Q^{ij}_{\rm mode} = \frac{2 \lambda_{2} M_2}{r^3}  \sum_m \mathcal{Y}_{lm}^{ij\ast} W_{lm} \kappa_{2m}  e^{-im\phi}, 
    \label{eq:Qij_in_eff_k2m}
\end{align}
where $\kappa_{2m}$ is given by Eq.~(\ref{eq:kappa_lm_from_mode}). To eliminate the dynamical component, we use the resummed equilibrium mode amplitude (cf. Eqs. \ref{eq:b_a_eq_N_w_im} and \ref{eq:b_a_eq_EOB_w_im})
\begin{align}
    b_a^{(\rm eq, PN)} \simeq \frac{\Delta_{a}^{(\rm PN)}\omega_{a0} \left[v_a + dv_a^{\rm (PN)}\right]}{\left[\Delta_a^{\rm (PN)}\right]^2 - i \left[m_a\omega + (2/3)(l_a+1)\Delta_a^{(\rm PN)} \right](\dot{\omega}/\omega)},
    \label{eq:b_a_eq_pn_w_im}
\end{align}
which reduces to Eq. (\ref{eq:b_a_eq_PN_nonres}) when the $\dot{\omega}/\omega$ term is ignored. Assuming $\omega_{a0}\to\infty$, we further have
\begin{align}
    \kappa_{2m}^{\rm (ad)} \simeq \frac{1}{z_{\rm PN}} + \frac{dv_a^{\rm (PN)}}{v_a}. 
    \label{eq:kappa_ad_limit}
\end{align}
This expression will be useful when verifying that our results reduce to those derived in \cite{Vines:11, Damour:12} under the adiabatic limit. 
Note that in this section, we use the same coordinate system (the harmonic coordinate system) as adopted in \cite{Vines:13}, which is different from the ones used for the EOB dynamics in Sec. \ref{sec:eob_hamiltonian}. This issue is circumvented by expressing our final result in terms of gauge-invariant variables $\omega$ and $v_\phi$ (which will be introduced later; see \cite{Damour:06}). 
For the mode amplitude, we use Eq. (\ref{eq:b_a_eq_pn_w_im}) expressed as a function of $\omega$ instead of directly using $b_a$ from the EOB evolution to separate the equilibrium component. 
In this coordinate system, we have
\begin{align}
    &\frac{dv_a^{\rm (pn)}}{v_a}|_{m_a=0} = \frac{6-X_2^2}{2}v^2 - \frac{5+3X_2}{2}\frac{M}{r} = \frac{1-3X_2 -X_2^2}{2}x \quad \text{ (circular orbit)}, \label{eq:dva_pn_m_0}\\
    &\frac{dv_a^{\rm (pn)}}{v_a}|_{|m_a|=2} = \frac{2-X_2^2}{2} v^2 - \frac{5 + 3 X_2}{2} \frac{M}{r} = \frac{-3-3X_2-X_2^2}{2} x \quad \text{ (circular orbit)}. \label{eq:dva_pn_m_2} 
\end{align}
where $v$ is the relative velocity and terms proportional to $\dot{r}$ are ignored. We have further taken the circular orbit limit in the last equality of each line. This result can be checked by matching $\kappa_{2m}^{\rm (ad)}$ to eq. (2.2b) of \cite{Vines:11} while using Eq. (\ref{eq:z_pn}) for $z_{\rm PN}$. 

Also needed when computing the system multiples are the derivatives of $Q^{ij}_{\rm mode}$ or equivalently derivatives of $q_a$ (as can be seen from Eq. \ref{eq:Qij}). Since these terms enter only at 1 PN, we can use the Newtonian mode dynamics (Eq. \ref{eq:dq_a_in}) to replace $\dot{q}_a$ by a function of $q_a$ (or $b_a$ since we work in the orbit frame). For this, we write
\begin{equation}
    [Q_{ij}^{\rm (ns)}]^{(k)}=\frac{2 \lambda_{20} M_2}{r^3}  \sum_m \mathcal{Y}_{lm}^{ij\ast} W_{lm} B_{2m}^{(k)}  e^{-im\phi}, 
\end{equation}
where
\begin{align}
    B_{2m}^{(1)} &= r^3 e^{im\phi}\frac{d}{dt}\left(\frac{\kappa_{2m}e^{-im\phi}}{r^3} \right), \nonumber \\
    &= -\frac{i}{2}\left[(\omega_{a0} + m(1+C_a)\Omega)\frac{b_{a,+}}{v_{a}} + (-\omega_{a0} + m(1+C_a)\Omega)\frac{b_{a,-}}{v_{a}}  \right], \\
    B_{2m}^{(2)} &= r^3 e^{im\phi}\frac{d^2}{dt^2}\left(\frac{\kappa_{2m}e^{-im\phi}}{r^3} \right), \nonumber \\
    &= -\frac{1}{2}\left[(\omega_{a0} + m(1+C_a)\Omega)^2\frac{b_{a,+}}{v_{a}} + (-\omega_{a0} + m(1+C_a)\Omega)^2 \frac{b_{a,-}}{v_{a}} - 2 \omega_{a0}^2 \right].
\end{align}
As in Eq. (\ref{eq:kappa_lm_from_mode}), here mode $a$ stands for the mode with $(l_a, m_a)=(2, m)$ and the sign following $a$ is the sign for $s_a$. 
A subtle point is that in the procedure above, we do not explicitly remove the dynamical component of the tide. Nonetheless, we note that if the equilibrium solution of Eq. (\ref{eq:b_a_eq_N_nonres}) is used in the final step, we have
\begin{align}
    B_{2m}^{(k)} = (-i m \omega)^k \kappa_{lm} \quad \text{ or }\quad Q_{{\rm mode}, lm}^{(k)} = (-i m \omega)^k Q_{{\rm mode}, lm},
    \label{eq:dBdt_eq}
\end{align}
for $k=1, 2$, which are the expected result. For the dynamical tide, taking a time derivative would lead to a prefactor $\propto [-i (\omega_a + m_a\Omega_1)]$ instead of the $(- i m \omega)$ we have above. 

Similarly, to 1 PN accuracy, we can use the conservative Newtonian dynamics to eliminate $\ddot{r}$ and $\ddot{\phi}$ for the orbit part in the intermediate steps as 
\begin{align}
    \ddot{r} \to r \dot{\phi}^2 - \frac{M}{r^2} + g_r, \quad\text{ and }\quad \ddot{\phi}\to - 2\frac{\dot{r} \dot{\phi}}{r}, 
\end{align}
where $g_r$ is the radial acceleration including both 1 PN and tidal contributions. 
Note that we do not include the tidal torque here as the lag in the tidal bulge is caused by the GW decay, which is turned off in this reduction process. This also means we will treat $H_{\rm mode}$ and $S_{1z, {\rm mode}}$ as constants during the intermediate steps. 

Using eqs. (3.31, 4.5a-c, 4.6, B4, B5) of \cite{Vines:13} with the derivative reduction procedure described above, we find the $(2,2)$ mode of the system mass moment as
\begin{align}
    Q_{\rm sys, 22}&=Q_{\rm mode, 22} + Q_{\rm pp, N, 22}
    \Bigg(
        1 + \left[\frac{1-3\eta}{6}v^2-\frac{5-8\eta}{7}\frac{M}{r}\right]  \nonumber \\
    & 
     + \frac{X_2 H_{\rm mode}}{(1-X_2)M} + \frac{4 X_2}{3(1-X_2)}\frac{\omega S_{1z, {\rm mode}}}{M} \nonumber \\
    &- \frac{X_2 \lambda_{20}}{56 (1-X_2)r^5}\Big\{[(6-104X_2+60X_2^2)\kappa_{20} + (513-484X_2+90X_2^2) \kappa_{22} + (9+36X_2+90X_2^2)\kappa_{2,-2}]\frac{M}{r} \nonumber \\
    &+ 4X_2(-11 \kappa_{20} + 52 \kappa_{22})r^2 \omega^2
     \Big\} 
    \Bigg),
\end{align}
where 
\begin{equation}
    Q_{\rm pp, N, 22} = \sqrt{\frac{15}{32 \pi}} \mu r^2 e^{-2i\phi},
\end{equation}
is the pp, Newtonian value. We have dropped terms $\propto \dot{r}$ and used Eq. (\ref{eq:dBdt_eq}) to replace the derivatives of the NS quadrupole assuming specifically the equilibrium tide. 

In order to use $Q_{\rm sys, 22}$ within the EOB code, we need to further eliminate $r$ in terms of the gauge-invariant $\omega$. If the orbit remains quasi-circular, the binary sits at the bottom of the radial effective potential with $\ddot{r}=0$, leading to 
\begin{equation}
    r^3 = \frac{M}{\omega^2} - \frac{r^2 g_r}{\omega^2}.
\end{equation}
We can then iteratively solve for $r(\omega)$ given $g_r$. The radial acceleration can be obtained from eqs. (5.9a)-(5.9e) of \cite{Vines:11} with 1 PN corrections to $Q_{\rm mode}^{ij}$ from their eq. (6.6b),  
\begin{align}
    g_r&=\frac{M}{r^2}\left[(-1-3\eta) v^2 + 2(2+\eta)\frac{M}{r}\right] 
    - \frac{M}{r^2}\frac{9X_2 M\lambda_{20} (2\kappa_{20}+3\kappa_{22}+3\kappa_{2,-2})}{8r^5 (1-X_2)} \nonumber \\
    -&\frac{M}{r^2}\frac{15 X_2 \lambda_{30} \left[5(\kappa_{33}+\kappa_{3,-3}) + 3(\kappa_{31}+\kappa_{3,-1})\right]}{4(1-X_2)r^7} \nonumber \\
    +&\frac{M}{r^2}\Biggl(-\frac{H_{\rm mode}}{M} + \frac{2+X_2}{1-X_2} \frac{\omega S_{1z,{\rm mode}}}{M} \nonumber \\
    +&\quad \frac{3X_2 \lambda_{20}}{8 r^5 (1-X_2)} \Bigg\{ 
        \frac{2M}{r}\Big[(16 + 18 X_2 -10 X_2^2)\kappa_{20} + (24+27X_2-15X_2^2)(\kappa_{22}+\kappa_{2,-2}) \Big] \nonumber \\
    &\quad\quad\quad\quad -3r^2\omega^2\Big[(2+4X_2 - 6X_2^2)\kappa_{20} + (3 + 8X_2 - 9X_2^2)(\kappa_{22} + \kappa_{2,-2}) \Big]
    \Bigg\}\nonumber \\
    +&\quad \frac{9 X_2(2-X_2)\lambda_{20 }}{4r^3 (1-X_2)} i \omega[B_{22}^{(1)} - B_{2,-2}^{(1)}] + \frac{27 X_2 \lambda_{20}}{16 r^3} [B_{22}^{(2)}+B_{2,-2}^{(2)}] \Biggr). 
\end{align}
The back-reaction from $l=3$ modes is included in the second line. The ratio of the Newtonian octupole tide to the 1-PN quadrupole tide is $(R/r)^2/x\simeq (R/M) (R/r)$ under the adiabatic limit, which means the octupole term becomes more important when $r\lesssim 3R$. The FF response of the $l=3$ f-mode makes its impact significant at a wider separation. 

We can now solve for the modified $r(\omega)$ relation for circular orbits as
\begin{align}
    r&=\frac{M^{1/3}}{\omega^{2/3}}\Bigg\{1 -\frac{(3-\eta)x}{3} + \frac{3 X_2  }{(1-X_2)} \frac{x^5\lambda_{20} }{M^5} \frac{2\kappa_{20}+3\kappa_{22}+3\kappa_{2,-2}}{8}\nonumber \\
    &+\frac{5 X_2}{(1-X_2)}\frac{x^7\lambda_{30}}{M^7}\frac{5(\kappa_{33}+\kappa_{3,-3})+3(\kappa_{31}+\kappa_{3,-1})}{4} \nonumber \\
    &+\frac{H_{\rm mode}}{3 M} - \frac{(2+X_2) x^{3/2} S_{1z,{\rm mode}}}{3(1-X_2)M^2}\nonumber \\
    &+\frac{X_2}{8 (1-X_2)} \frac{x^6 \lambda_{20}}{M^5} [(24-22 X_2) \kappa_{20} + (36-33X_2)(\kappa_{22} + \kappa_{2,-2})]
    \Bigg\}, 
\end{align}
where Eq. (\ref{eq:dBdt_eq}) has been used for the equilibrium tide. 
When the adiabatic limit of $\kappa_{2m}$ is taken (including PN corrections; Eq. \ref{eq:kappa_ad_limit}), the result reduces to eq. (2.9) of \cite{Vines:11} and the tidal spin-orbit term matches its BBH counterpart (e.g., eq. 16 of \cite{Kidder:93}). 

Plugging $r(\omega)$ into $Q_{\rm sys, 22}$ and using the equilibrium relations, we arrive at our final result
\begin{align}
    Q_{\rm sys, 22} = \sqrt{\frac{15}{32\pi}} \frac{\eta M^{5/3}}{\omega^{4/3}}e^{-2i\phi} \left[1 - \frac{107-55\eta}{42} x + \frac{h_{22}^{(t)}}{h_{22}^{'(N,0)}} \right],
\end{align}
where the fractional correction due to tide is given by
\begin{align}
    \frac{h_{22}^{(t)}}{h_{22}^{\rm '(N,0)}} &= 
    \frac{3}{(1-X_2)} \frac{x^5\lambda_{20}}{M^5}\left[\kappa_{22} + \frac{X_2(2\kappa_{20}+3\kappa_{22}+3\kappa_{2,-2})}{4}\right]  \nonumber \\
    &+\frac{5X_2}{(1-X_2)}\frac{x^7 \lambda_{30}}{M^7}\frac{5(\kappa_{33}+\kappa_{3,-3})+3(\kappa_{31}+\kappa_{3,-1})}{2}  \nonumber \\
    &+\frac{2+X_2}{3(1-X_2)}\frac{H_{\rm mode}}{M} - \frac{2(2-X_2)}{3(1-X_2)}\frac{x^{3/2} S_{1z,{\rm mode}}}{M^2} \nonumber \\
    &+\frac{1}{28 (1-X_2)}\frac{x^6 \lambda_{20}}{M^5}
    \Big[ 2X_2(75 - 42X_2 - 13X_2^2)\kappa_{20} \nonumber \\
    &\hspace{3cm} + 3(84 - 51X_2 - 5 X_2^2 - 13 X_2^3) \kappa_{22} 
        +3X_2(61 - 85 X_2 -13 X_2^2) \kappa_{2,-2}
    \Big].
    \label{eq:frac_h_22_t}
\end{align}
The fractional correction to $Q_{\rm sys, 22}$ is the same as the one to the $(2,2)$ GW mode, which follows directly from Eq. (\ref{eq:h_lm_from_multipoles}) together with the equilibrium condition $d^2 Q_{\rm sys, 22}/dt^2 = - 4 \omega^2 Q_{\rm sys, 22}$ (Eq. \ref{eq:dBdt_eq}).
We check that in the adiabatic limit, the expression in Eq. (\ref{eq:frac_h_22_t}) reduces to eq. (3.3a) of \cite{Vines:11} and eq. (A14) of \cite{Damour:12}. 
We also remind the readers that $\kappa_{lm}$ is computed using Eq. (\ref{eq:kappa_lm_from_mode}) with mode amplitudes given by Eq.~(\ref{eq:b_a_eq_pn_w_im}) and treated as a function of $\omega$. To further compute $b_a^{\rm (eq, PN)}/v_a$, we use $\Delta_a^{\rm (PN)}$ from Eq. (\ref{eq:Delta_a_PN}) with $z_{\rm PN}$ and $\Omega_{\rm FD}$ given respectively by Eqs. (\ref{eq:z_pn}) and (\ref{eq:Omega_FD}), and $dv_a^{\rm (PN)}/v_a$ from Eqs. (\ref{eq:dva_pn_m_0}) and (\ref{eq:dva_pn_m_2}). The expression is supplemented with the leading-order relation $\dot{\omega}/\omega^2\simeq (96/5) \eta x^{5/3}$. Lastly, the $H_{\rm mode}$ and $S_{1z,{\rm mode}}$ terms appear only in the PN corrections, so we simply take their values from the EOB evolution. Note these terms contain the full tidal contribution (including both equilibrium and dynamical tides) as they take the form $\propto b_a^\ast b_a$ instead of $\propto b_a^\ast v_a$ and thus vary only on the GW decay timescale. 
The companion's tidal contribution is obtained by swapping $X_2$ with $X_1$. 

In analogy to the treatment of radial interaction, we can define another effective Love number, $\kappa_{{\rm eff}, h}$, that preserves the form of the adiabatic tide result derived in eq. (A14) of \cite{Damour:12} and captures finite frequency corrections through a replacement $\lambda_2 \to \kappa_{{\rm eff}, h}\lambda_2$. At the Newtonian order and focusing on the $l=2$ tide, this means 
\begin{equation}
    (1+2X_2)\kappa_{{\rm eff}, h} = \kappa_{22} + \frac{X_2(2\kappa_{20} + 3\kappa_{22} + 3\kappa_{2,-2})}{4} = \kappa_{22} + 2 X_2 \kappa_{{\rm eff},r}, 
\end{equation}
where $\kappa_{{\rm eff}, r}$ is the effective Love number that preserves the radial interaction (Eq. \ref{eq:kappa_eff_r})
\begin{align}
    \kappa_{{\rm eff}, r} = \frac{2 \kappa_{20} + 3 \kappa_{2,2} + 3\kappa_{2,-2}}{8} \simeq \frac{1}{4} + \frac{3}{4} \kappa_{22}, 
\end{align}
where the last equality follows when the imaginary components in $\kappa_{2,\pm2}$ are ignored. 
We therefore have
\begin{equation}
    \kappa_{{\rm eff}, h} \simeq \frac{-1 + 4 \kappa_{{\rm eff}, r} + 6 X_2\kappa_{{\rm eff}, r} }{3 (1+2 X_2)},
    \label{eq:kappa_eff_h_vs_kappa_eff_r}
\end{equation}
as a relation connecting the two effective Love numbers that separately preserve the radial interaction and GW radiation at the Newtonian order. This appears to be different from the one implemented in \texttt{LAL} \cite{LAL:18} as described in Footnote \ref{ft:kappa_h_in_lal}.

For 1 PN accuracy, we also need tidal corrections to $Q_{{\rm sys}, 3m}$ and $S_{{\rm sys}, 21}$ to the Newtonian order. Under the circular orbit limit, the fractional corrections to them (which are the same as fractional corrections to the GW modes they source) are
\begin{align}
    \frac{h^{(t)}_{33}}{h^{\rm '(N, 0)}_{33}} &= - \frac{x^5 \lambda_{20}}{M^5} \frac{9X_2  \left[(-2+4X_2)\kappa_{20} + (5+6X_2)\kappa_{22} + (-3+6X_2)\kappa_{2,-2}\right] }{8(1-X_2)} , \\
    \frac{h^{(t)}_{31}}{h^{\rm '(N, 0)}_{31}} &= - \frac{x^5 \lambda_{20}}{M^5 }\frac{3X_2 \left[(10 + 12 X_2) \kappa_{20} + (-1+18X_2)\kappa_{22} + (-9 + 18 X_2)\kappa_{2, -2}\right]}{8 (1-X_2)}, \\
    \frac{h^{(t)}_{21}}{h^{\rm '(N, 1)}_{21}} &= -\frac{x^5 \lambda_{20}}{M^5}\frac{3X_2\left[(-2+12X_2)\kappa_{20}+(-1+18X_2)\kappa_{22}+(-9+18X_2)\kappa_{2,-2}\right]}{8 (1-X_2)}. 
\end{align}
The tidal contribution to the system octupole mainly comes from the $l=2$ tide's back-reaction, whose effect is greater than the NS octopole $Q_{\rm mode, 33}$ by a factor of $(R/r)^2$. Similarly, we ignore $S^{ij}_{\rm mode}$ when computing the current quadrupole. The adiabatic limit matches eqs. (3.4) and (3.5) of \cite{Vines:11}, and eqs. (A16) and (A17) of \cite{Damour:12}. If the companion is also an NS, its tidal contribution is obtained by swapping $X_2$ with $X_1$ and flipping the overall sign. In other words, whereas the tidal effects add in the leading, $(2,2)$ GW mode, their effects subtract in the next order $(3, m)$ and $(2, 1)$ GW modes, and the $(3, m)$ and $(2, 1)$ GW modes vanish if two NSs are identical. 

Once $h_{lm}^{(t)}$ is obtained, we follow \cite{Damour:12} and treat them as an additive modification to the total GW waveform, 
\begin{equation}
    h_{lm} = h^{\rm (pp)}_{lm} + h^{(t)}_{lm}, 
\end{equation}
where the pp waveform $h^{\rm (pp)}$ we take from \cite{Pan:11} (see their appendix B for details); their leading order Newtonian expression is given in Eq. (\ref{eq:h_lm_pp_N}). We also replace $x=\hat{\omega}^{2/3}$ appearing in $h^{(t)}_{22}$ by $v_\phi^2$. 
The energy loss rate, in terms of the GW modes, is given by (\cite{Pan:11}; note we use a different sign convention)
\begin{equation}
    \frac{d\hat{E}}{d\hat{t}} = - \frac{\hat{\omega}^2}{8\pi}\sum_{l}\sum_{m>0} m^2 \Big{|}\frac{D_L}{M}h_{lm}\Big{|}^2,
\end{equation}
and the GW induced torque, $\hat{\mathcal{F}}_\phi$ (see Eq. \ref{eq:dP_phi_dt}), is given in terms of the energy loss
\begin{equation}
    \hat{\mathcal{F}}_\phi = \frac{\mathcal{F}_\phi}{\mu M} = \frac{1}{\eta \hat{\omega}}\frac{d\hat{E}}{d\hat{t}}.  
\end{equation}
In our calculation, we include the pp part of the $(2, 2)$, $(2, 1)$, $(3, 3)$, $(3, 1)$, $(3, 2)$, $(4, 4)$ GW modes and tidal corrections are added to the first 4 modes to 1 PN accuracy. The higher PN corrections to the tide including FF responses are recently derived in \cite{Mandal:23, Mandal:24} yet we find the 1 PN effects are already small (the orbital dynamics is mainly dominated by the conservative tidal torque near f-mode resonance as shown in \cite{Yu:24a}), and defer their implementation to future work.

\section{Results}
\label{sec:results}

Having described the conservative (Sec. \ref{sec:cons_dyn}) and dissipative (Sec. \ref{sec:GW_rad}) tidal effects, we adapt them into the spin EOB waveform (specifically, the \texttt{SEOBNR} family available in \texttt{LAL} \cite{LAL:18}), similar to the construction done by \cite{Steinhoff:21}. In particular, we replace the pp part in Eq. (\ref{eq:H_in_eff}) with those given in \cite{Barausse:10} (see also \cite{Barausse:11, Taracchini:12}). This includes replacing the potentials $A_{\rm pp}$ and $D_{\rm pp}$ and including the spin part of the Hamiltonian (i.e., the $H_{\rm S}$ and $-\mu S_\ast^2/2Mr^3$ terms in the notation of \cite{Barausse:10}). As we have explicitly included the interactions due to the tidal spin, we use the constant background spins ($\chi_1$ and $\chi_2$) when evaluating the Kerr parameter and other spin-related terms in $H_{\rm S}$. We use $p_\phi$ whenever the orbital angular momentum appears in $H_{\rm S}$. 
The (equilibrium) tidal contribution to the GW strain is added to the pp part described in \cite{Pan:11, Taracchini:12}. We ignore the dynamical tide contribution to the strain that varies at $\omega_a + m_a \Omega_{1}$, which does not couple with the orbital quadrupole (Eq. \ref{eq:dEdt_vs_quad}) and hence does not produce $(R_1/r)^5$ correction to the problem. 
Our code allows the companion to be either an NS or a BH. For the numerical results presented in this work, we will fix the companion to be an NS with $M_2=M_1=1.35\,M_\odot$ and $\chi_2=0$. 

We adopt the SLy EOS \cite{Douchin:01} for the NS. For a typical NS with mass $M_1=1.35\,M_\odot$, we adopt $R_1=11.7\,{\rm km}$, $\lambda_{2}/M_1^5=389$, $\lambda_{3}/M_1^7=700$, $M_1 \omega_{a0}|_{l_a=2}=0.07934$, $M_1 \omega_{a0}|_{l_a=3}=0.1067$, $I_1/M_1^3=12.34$, and $C_a|_{l_a=2}=-1/4$; see table I of \cite{Steinhoff:21} and their eq. (5.7). A more complicated fit of $C_a$ is also provided in eq. (5.5) of \cite{Steinhoff:21}, yet we caution the use of it as it was obtained under the Cowling approximation, which may be inaccurate for f-modes \cite{Ranjan:24}. Alternatively, the fit in \cite{Kruger:20} may be used, yet their result is most readily adopted when one fixes the NS's central energy density while we fix its mass. A user of the code has the freedom to choose the value of $C_a$.  We also take $C_{ES^2}=6$ following the notation of \cite{Steinhoff:21}, which is a coefficient describing the spin-induced quadrupole and is normalized to 1 for BHs \cite{Poisson:98, Krishnendu:17}. 

\subsection{Waveforms}

\begin{figure}
    \centering
    \includegraphics[width=0.75\linewidth]{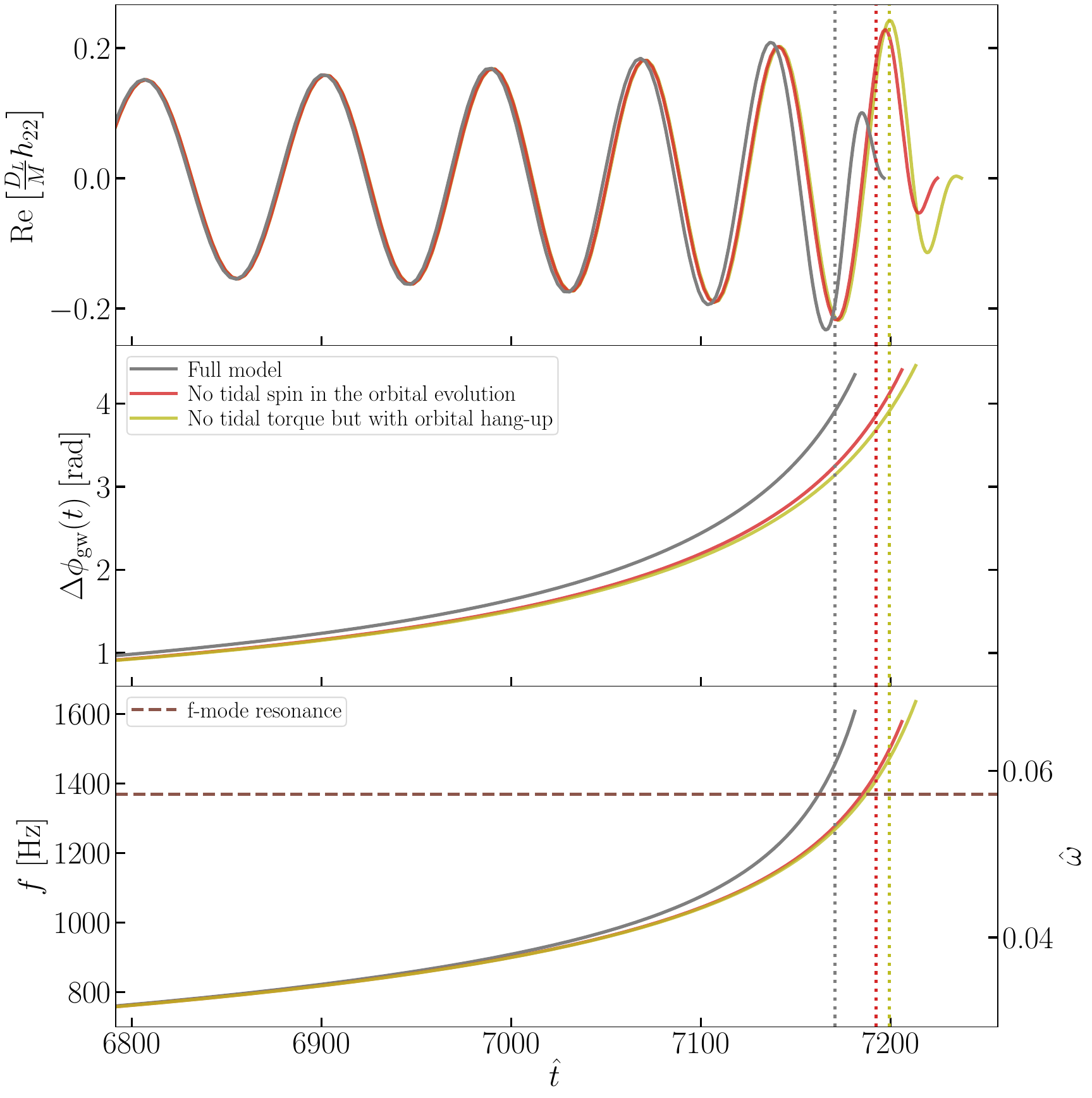}
    \caption{Same as the system shown in the lower panel of Fig.~\ref{fig:h22_tidal_spin} ($\chi_{1z}=-0.2$) but also includes each waveform's phase shift relative to the pp waveform (middle panel) and the time-frequency trajectory (bottom panel). Also shown in the bottom panel is the resonance frequency of the $l_a=m_a=2$ f-mode, which is reached slightly before the waveform's amplitude peaks. At the peak amplitude, the phase difference between the models with and without the tidal spin is about 0.6 rad. }
    \label{fig:full_ts_vs_nts_modspin}
\end{figure}

Our key result has been presented in Fig.~\ref{fig:h22_tidal_spin} where we show the dominant (2, 2) GW mode as a function of time near the merger. The difference between the full model (shown in gray) and the one without the tidal spin (shown in red; this is essentially the effective Love number model described in \cite{Steinhoff:21}) is remarkably similar to the difference between numerical relativity and the model of \cite{Steinhoff:21} as presented in the top panel of their fig. 3.\footnote{The adiabatic Love number we adopted to generate Fig.~\ref{fig:h22_tidal_spin} assumes the SLy EOS and is about half of the one used in \cite{Steinhoff:21} for their fig. 3. This is compensated by the fact that we used a BNS system whereas \cite{Steinhoff:21} assumed an NSBH. } This suggests that the tidal spin is a crucial component required in the construction of faithful analytical BNS and NSBH models all the way to the final merger when the (2, 2) GW mode reaches its largest amplitude (see also the comparison of different terms in the Hamiltonian in Fig. \ref{fig:H_comp}).

The tidal spin is especially important when the NS has an anti-aligned background spin $\chi_{1z}<0$ as demonstrated in the lower panel of Fig. \ref{fig:h22_tidal_spin}. A closer examination of this system with $\chi_{1z}=-0.2$ is presented in Fig. \ref{fig:full_ts_vs_nts_modspin}. This is also the same system we used to generate Figs. \ref{fig:bulge}-\ref{fig:H_comp}. From top to bottom, we show the GW strain, the dephasing relative to a pp waveform (with $\Delta \phi_{\rm gw}(t)=2[\phi(t) - \phi_{\rm pp}(t)]$), and the $f-t$ trajectory with $f=\omega/\pi$ the GW frequency. The vertical dotted lines indicate locations where the (2, 2) GW mode reaches its maximum amplitude, and in the middle and bottom panels, the last point in each curve is where $r=R_1+R_2$. 

We still use the color gray to represent the full model and red for the case where we set $S_{1z, {\rm mode}}=S_{2z, {\rm mode}}=0$ when evolving the orbital variables. When evolving the modes, we keep the tidal spins in the red curves to capture the frame-dragging effects in mode resonance. This way, the model represented in red is essentially the EOB model based on the effective Love number approach described in \cite{Steinhoff:21}.\footnote{There are a few minor differences. For example, we keep the PN corrections to the mode resonance while \cite{Steinhoff:21} adopted the Newtonian approximation. The difference between the two is small as described in Fig. \ref{fig:GR_freq_shift}. We also have different treatments in the tidal correction to the GW radiation as described in Sec. \ref{sec:GW_rad}. Nonetheless, \cite{Yu:24a} showed that the conservative tidal effect dominates near mode resonance. } In addition, we also present a model in the yellow curve where we set $\dot{p}_\phi = \mathcal{F}_\phi$ (instead of $\dot{P}_\phi = \dot{p}_\phi + \dot{S}_{1z, {\rm mode}}=\mathcal{F}_\phi$). In other words, we eliminate the Newtonian tidal torque acting on the orbit but still allow the tidal spin to correct the dynamics through PN effects (especially the orbital hang-up \cite{Campanelli:06}). In Fig. \ref{fig:full_ts_vs_nts_modspin}, the dominant effect is the Newtonian tidal torque (comparing the yellow and gray models), which accelerates the inspiral, causing the merger to happen $29M$ earlier compared to the case where it is ignored. On the other hand, since the tidal spin is always positive (in the direction of the orbital angular momentum), the PN orbital hang-up it creates will counteract the tidal torque and cause the inspiral to last slightly longer and reach a higher frequency at a fixed separation (comparing the red and yellow models). Summing both effects together, the full model (gray) merges $22 M$ before the effective Love number model (red) and the phase difference between the two is 0.7 rad at the (2,2) GW mode's peak amplitude.


\begin{figure}
    \centering
    \includegraphics[width=0.75\linewidth]{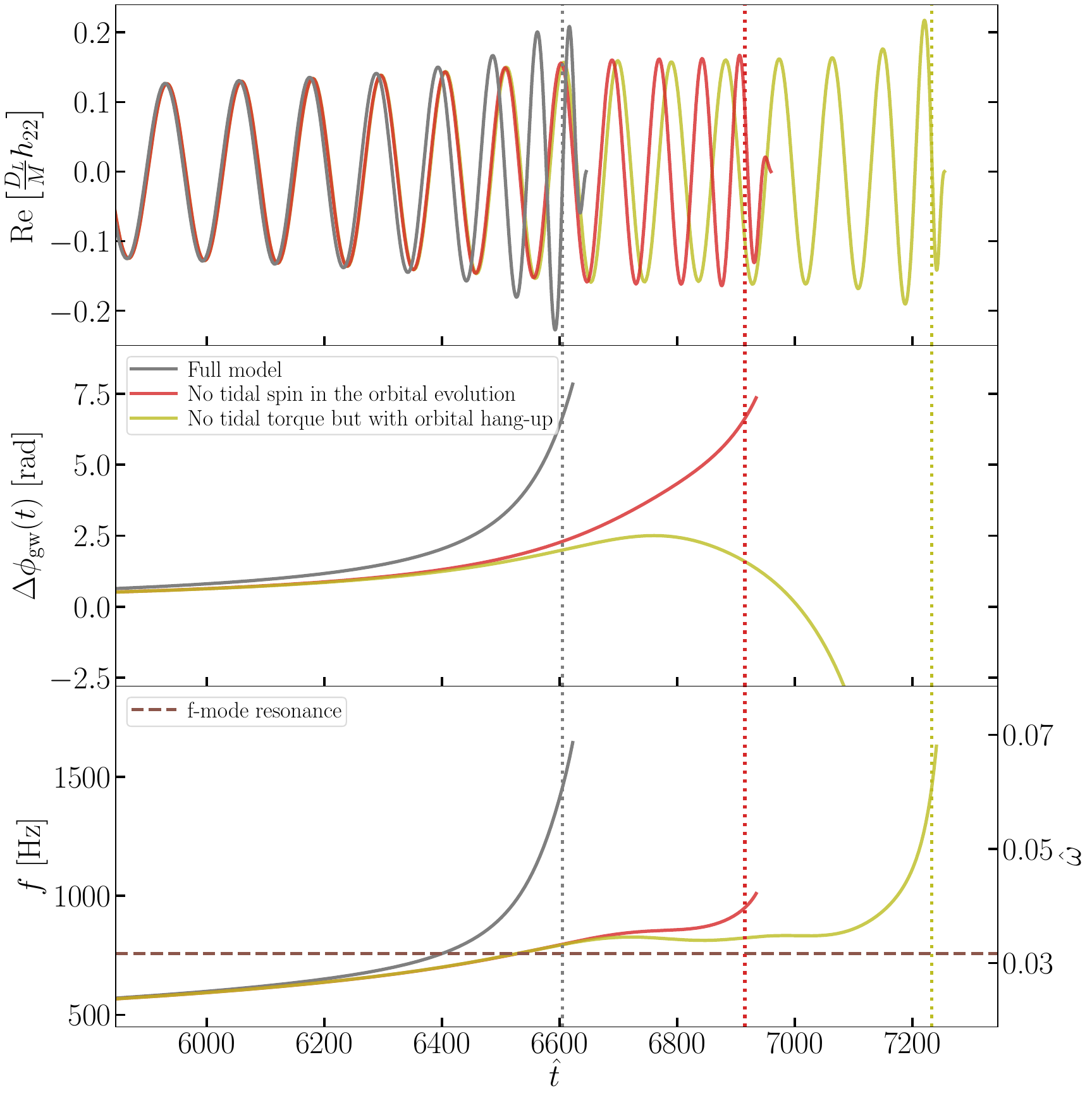}
    \caption{Similar to Fig.~\ref{fig:full_ts_vs_nts_modspin} but for a BNS where one NS has a more rapid spin ($\chi_{1z}=-0.4$, about half of its break-up frequency). The other NS is still non-spinning. 
    The pp waveform peaks at $\hat{t}=7305$ and the tide shortens the inspiral by $\Delta \hat{t}_{\rm pk}=512$. 
    Note that ignoring the tidal spin completely (as in the red curves) ignores both the back-reaction torque (mostly Newtonian) and the PN orbital hang-up due to the tidal spin-orbit interaction. The case where only the tidal torque is disabled but the orbital hang-up is still active is shown in the olive curves. 
    Whereas the tidal torque shortens the inspiral (comparing gray and olive), it is partially compensated by the orbital hang-up effect which increases the duration of the inspiral (comparing olive and red curves). Things are more complicated because the excited f-mode can oscillate at its own frequency and becomes phase-incoherent with the orbit. The excited f-mode can then transfer some energy back to the orbit and drive it to be eccentric (as can be seen in the non-monotonic frequency evolution in the bottom panel). 
    The net phase shift caused by the tidal spin is about 1.4 rad at the peak of $|h_{22}|$. 
    } 
    \label{fig:full_ts_vs_nts_highspin}
\end{figure}

We consider an even more rapidly rotating NS in Fig. \ref{fig:full_ts_vs_nts_highspin} where the background spin is $\chi_{1z}=-0.4$, or in physical units, $\Omega_{1}/2\pi = -780\,{\rm Hz}$ (for the specific EoS we considered). This is similar to the spin rate of the fastest-spinning pulsar known to date and is about half of the breakup frequency of the chosen EoS \cite{Read:09a}. This configuration leads to an early excitation of the f-mode at $f=757\,{\rm Hz}$ and consequently a strong dynamical tide. The tidal spin can reach $\chi_{1z, {\rm mode}}\simeq 0.39$, almost as large as the background spin (Fig. \ref{fig:chi1_t}). As a result, both the Newtonian torque and the PN orbital hang-up can have significant dynamical impacts. The Newtonian torque shortens the inspiral by almost 7 cycles while the PN hang-up undoes half of that. It is also interesting to note that the full model ends at a much higher frequency when $r=(R_1+R_2)$ compared to the red model, because the tidal spin cancels largely with the background, making the total net spin of the NS less negative. 

\subsection{Deviation from the circular approximation}
\label{sec:ecc}

\begin{figure}
    \centering
    \includegraphics[width=0.75\linewidth]{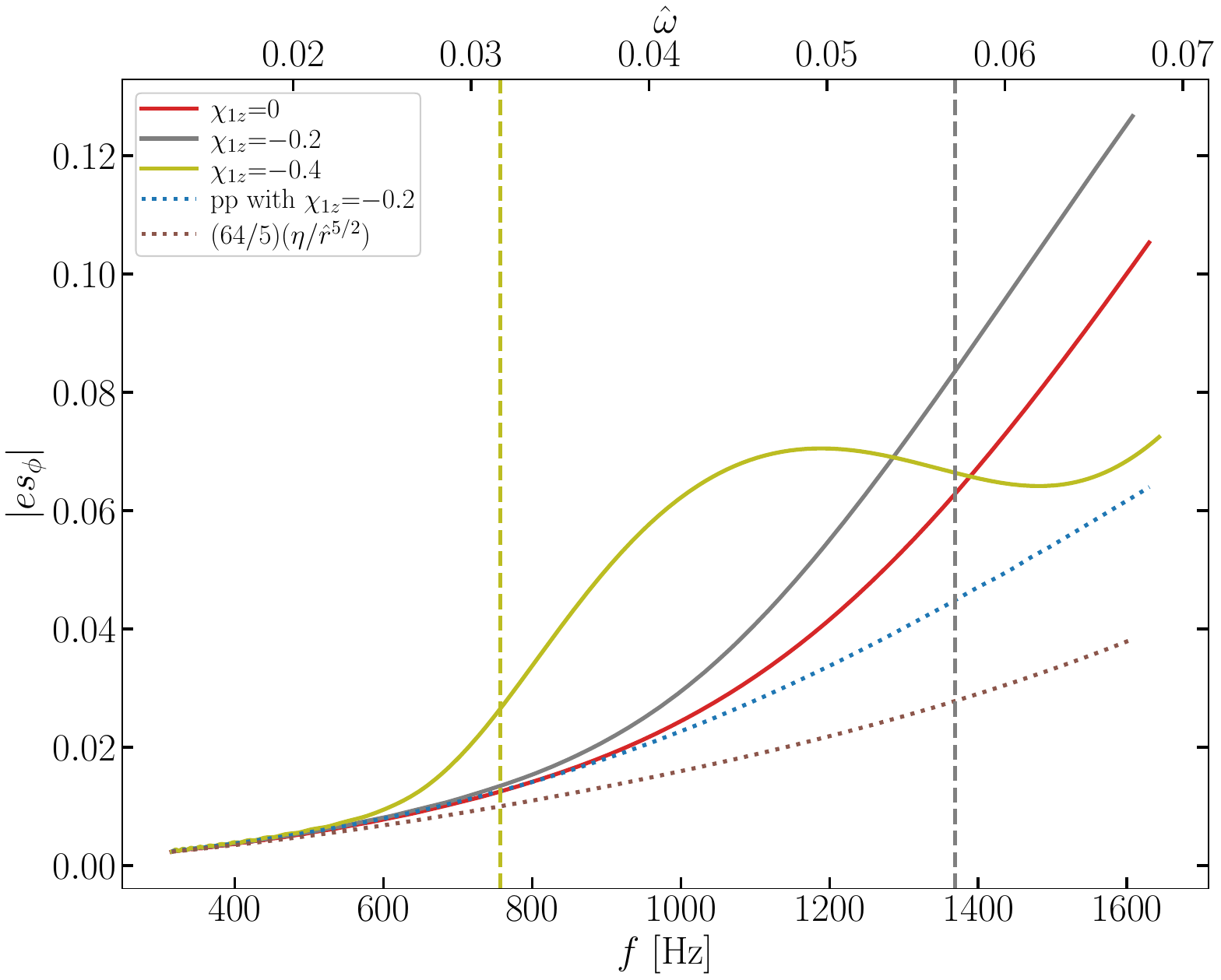}
    \caption{Osculating eccentricity $e s_\phi$.  
    } 
    \label{fig:esp}
\end{figure}

As a caveat, we note that once the f-mode is resonantly excited, the orbit cannot remain quasi-circular as demonstrated in \cite{Yu:24a}. On the other hand, our EOB model assumes quasi-circular orbits when computing both the conservative Hamiltonian and the radiative multipoles.

To assess the deviation from the circular approximation, we consider the orbital eccentricity evolution during the inspiral. Defining eccentricities for a relativistic orbit can be subtle, yet we can use a gauge-dependent osculating definition as an approximation. At any moment, we can map $(r, \dot{r}, \phi, \dot{\phi}=\omega)$ to $(p_{\rm slr}, e, \phi, \phi_0)$, which are respectively the instantaneous semi-latus rectum, eccentricity, orbital phase, and argument of pericenter. Of particular interest are~\cite{Yu:24a} 
\begin{equation}
    p_{\rm slr} = \frac{r^4 \omega^2}{M}, \quad \text{ and } e s_\phi= \dot{r} \sqrt{\frac{p_{\rm slr}}{M}}, 
\end{equation}
where $s_\phi = \sin (\phi - \phi_0)$. For a quasi-circular pp orbit at the Newtonian order, we have (eq. 108 of \cite{Yu:24a})
\begin{equation}
    |e s_\phi|_{\rm pp} \simeq \frac{64}{5} \eta \left(\frac{M}{r}\right)^{5/2},  
    \label{eq:esp_pp_N}
\end{equation}
due to the gradual decay of the orbit. Near a mode's resonance, the tidal torque efficiently transfers energy and AM from the orbit to the mode, making $|\dot{r}|$ and consequently $|e s_\phi|$ greater. Post resonance, $e s_\phi$ will oscillate at a frequency $(\omega_a + m_a\Omega_1 - m_a \omega)$ due to the beat between NS and orbital quadrupoles. The $ec_\phi$ term (due to radial interactions) is small in comparison and ignored here. A detailed theoretical discussion on the eccentricity excitation can be found in sec. IV. C of \cite{Yu:24a}. 

We show in solid lines in Fig. \ref{fig:esp} the osculating eccentricity extracted from our EOB model for NSs with different values of background spin. We use the color (red, gray, yellow) for $\chi_{1z}=(0, -0.2, -0.4)$. Also shown in the blue-dotted line is $|es_\phi|$ extracted from the pp orbit with $\chi_{1z}=-0.2$, and in the brown-dotted line the value computed from Eq. (\ref{eq:esp_pp_N}). As the NS's background spin changes from $\chi_{1z}=0$ (red) to $\chi_{1z}=-0.2$ (gray), the eccentricity increases because the mode is closer to resonance and the tidal response is amplified through the Lorentzian (e.g., Eq. {\ref{eq:b_a_eq_N_nonres}}). Note that when resonance happens (indicated by the vertical-dashed lines in the plot), $e s_\phi \simeq 2 (e s_\phi)_{\rm pp}$. Indeed, from eqs. (122), (23), (55), and (70) of \cite{Yu:24a}, we have
\begin{align}
    \left[\frac{e s_\phi}{(e s_\phi)_{\rm pp}}\right]_{\rm res} &\simeq \frac{15}{2048}\sqrt{\frac{15\pi}{2}} \left(\frac{1}{4\eta}\right)^{3/2}(2X_1)^5\frac{X_2}{X_1}  \left(\frac{\lambda_2}{M_1^5}\right) (M_1\omega_{a0}) (M \omega_{\rm res})^{-1/6}, \nonumber \\
    &\simeq 1.2 \times \left(\frac{1}{4\eta}\right)^{3/2}(2X_1)^5\frac{X_2}{X_1}  \left(\frac{\lambda_2/M_1^5}{390}\right) \left(\frac{M_1\omega_{a0}}{0.08}\right) \left(\frac{M \omega_{\rm res}}{0.08}\right)^{-1/6},
\end{align}
which is approximately unity and only weakly depends on the orbital frequency of resonance, $\propto \omega_{\rm res}^{-1/6}$. Since $[(e s_\phi)_{\rm pp}]_{\rm res}$ decreases with $\omega_{\rm res}$ (Eq. \ref{eq:esp_pp_N}), this explains why once mode resonance happens during the inspiral, making $\chi_{1z}$ more negative and $\omega_{\rm res}$ smaller will reduce the eccentricity excitation caused by the tide (e.g., from gray to yellow).  This further suggests that for the entire range of allowed NS spin, the eccentricity is bounded with $|e s_\phi|< 0.12$. 

We note that the recent NR simulation of a rapidly rotating BNS in \cite{Kuan:24} showed no apparent deviation from quasi-circular approximation. This is consistent with our analysis above, and the time-frequency trajectory shown in the bottom panel of Fig. \ref{fig:full_ts_vs_nts_highspin} in the gray line. This differs from our previous analysis in \cite{Yu:24a} under the Newtonian limit. We attribute the difference to the PN acceleration of the inspiral due to anti-aligned background spin (comparing the blue-dotted and brown-dotted lines in Fig. \ref{fig:esp}). This makes the tidal back-reaction insufficient to change the sign of $\dot{r}$ in the full EOB model at the spin levels considered. The reduction in the post-resonance time also explains the lack of radiation at the f-mode frequency in \cite{Kuan:24}. 
Nonetheless, we note that when an NS has a more rapid anti-aligned spin with $\chi_{1z}\lesssim -0.5$, $\dot{r}$ can become positive because of the dynamical tide's back-reaction, making $r$ and $\omega$ non-monotonic functions of time. Similar effect is also observed in the inspiral of initially eccentric BNSs \cite{Takatsy:24}. 
While this should not be an issue for time-domain waveform models (like the one constructed here), it requires caution when constructing frequency-domain approximants.

\section{Conclusion and discussion}
\label{sec:discussion}

In this work, we presented a new EOB model that includes the NS dynamical tide. The source code of our model, which is publicly available at \url{https://github.com/hangyu45/EOBnsmodes}, can generate time-domain waveforms in both decomposed modes $h_{lm}$ (which we showed in Figs. \ref{fig:h22_tidal_spin}, \ref{fig:full_ts_vs_nts_modspin}, and \ref{fig:full_ts_vs_nts_highspin} for the dominant 22 mode) and two polarizations ($h_+$ and $h_\times$; e.g., eq. 71 of \cite{Blanchet:14}). 

Compared to previous models based on the effective Love number approach \cite{Steinhoff:16, Hinderer:16, Steinhoff:21}, the most significant extension made in our new model is the incorporation of the tidal spin (Fig. \ref{fig:chi1_t}) and its back-reaction to the orbit via both Newtonian torque (Sec. \ref{sec:tidal_spin}) and PN orbital hang-up (Eqs. \ref{eq:H_t_LS_eff}). Additionally, we self-consistently included mode resonance in the system multiple moments when computing the GW radiation (Sec. \ref{sec:GW_rad}), yet consistent with the analysis in \cite{Yu:24a}, we found the dissipative effect is subdominant compared to the conservative ones due to the tidal spin when the dynamical tide is excited. 

We showed in Fig. \ref{fig:h22_tidal_spin} that the difference between models with and without the tidal spin closely resembles the difference between NR and the previous EOB model developed in \cite{Steinhoff:21} (which is essentially the model without the tidal spin). This, at least based on a qualitative visual inspection, underscores tidal spin as a necessary component to analytically explain results from NR.
We defer a direct comparison and calibration of our current model to NR\footnote{In private communication, colleagues from the Max Planck Institute for Gravitational Physics \cite{private_comm_Haberland} kindly shared a direct comparison between our model with the NR simulation of \cite{Kuan:24}, which showed improvement of our model over previous ones \cite{Steinhoff:21, Gamba:23b}, especially in the resonance and post-resonance regimes. } as the nonlinear hydrodynamical mode interactions studied in \cite{Yu:23a} should be included in the model first to complete the analytical description. \cite{Yu:23a} demonstrated that the nonlinear interaction is significant at the Newtonian order as it effectively shifts the f-mode natural frequency and enhances the mode's resonance. Interestingly, the f-mode frequencies used in \cite{Steinhoff:21} to match NR do not follow the universal relation with their associated Love numbers. The deviations from the universal relation are likely caused by the negligence of the nonlinear hydrodynamics, which would cause a shifted frequency instead of the intrinsic, linear one (which follows the universal relation) to be used. 
We plan to explore this in a follow-up study, which should also include a systematic validation and calibration against NR.  

Our model so far considers the NS f-mode only. Nonetheless, it is set up to be able to include an arbitrary set of electric-type modes (including gravity \cite{Lai:94c, Reisenegger:94,  Yu:17a, Yu:17b, Kuan:21, Kuan:21b, Counsell:25}, pressure \cite{Andersson:20, Passamonti:22} and interface \cite{Tsang:12, Pan:20, Passamonti:21} modes in addition to the fundamental ones; see \cite{Suvorov:24} for a recent review) as we evolve each mode's amplitude instead of a single $Q^{ij}$ summed over all the modes. 
Recently, \cite{Kwon:24} showed that the NS g-mode can cause a large, nearly 3 rad phase shift to the waveform because it can evolve into a resonance-locking state thanks to nonlinear mode interactions that correct its effective frequency as the orbit evovles. However, \cite{Kwon:24} considered the effect only at the Newtonian order. Because this effect happens as early as $f\simeq100\,{\rm Hz}$, it is hard to be simulated by NR. Therefore, an analytical model involving nonlinear hydrodynamics, like the one we propose to study above, would be ideal to address the effects of g-mode resonance locking in a relativistic context. Likewise, our model can also be extended to study modes excited by the gravitomagnetic PN potential that are resonantly excited in the early part of the inspiral \cite{Ho:99, Flanagan:07, Poisson:20, Gupta:21, Ma:21, Mandal:23}. 

Another future direction is to consider the impact of the tidal spin in the context of spin precession \cite{Buonanno:06, Ossokine:20, Akcay:21, RamosBuades:23, Yu:23b}. Since the dynamical tide is most significant in rapidly rotating NSs with anti-aligned spins, such an NS is likely assembled into a binary through dynamical formation \cite{Rodriguez:16}. In this case, the (background) spin axis would in general be randomly aligned with the orbital AM, leading to precessions of both vectors. The tidal spin enters the PN dynamics in exactly the same way as the background spin \cite{Steinhoff:16, Mandal:24} except that its magnitude evolves over time, which can lead to interesting effects in the precession dynamics. It is well known that the effective spin ($=M_1 \chi_{1z} + M_2\chi_{2z}$), while remaining constant in BBH systems \cite{Racine:08}, is not conserved by the precession in BNS and NSBH systems \cite{LaHaye:23}. Our study further suggests that the net spin magnitude of each NS will also evolve, providing a further distinction from the BH case.   

The dynamical formation of a binary is also likely to produce some orbital eccentricity. This is also a crucial component to be included in future studies. As an eccentric orbit has higher frequency harmonics, it allows the NS dynamical tide to be excited without requiring a particularly high spin rate \cite{Gold:12, Chirenti:17, Yang:18, Vick:19b, Parisi:18, Yang:19, Wang:20, Takatsy:24}. Meanwhile, once the dynamical tide is excited, it will inevitably force the orbit to be eccentric (Sec. \ref{sec:ecc}). While we argued in Sec. \ref{sec:ecc} that the osculating eccentricity due to the dynamical tide is bounded to within 0.1, the error it caused in the waveform needs to be quantified with a model capable to describe at least a moderately eccentric orbit. For this, integrating matter effects into a relativistic model allowing orbital eccentricity \cite{Hinderer:17, Khalil:21, Nagar:21, Liu:22} would be crucial. 

On the fundamental physics side, our new EOB model opens the possibility to test general relativity via the $I$-Love universal relation \cite{Yagi:13} with GW observations alone. 
\cite{Yagi:13} showed that as long as relativity holds, the NS moment of inertial $I$ and its Love number $\lambda$ will follow a nearly universal, EoS-independent relation. On the other hand, alternative theories of gravity lead to deviations from this relation. 
In the original proposal, the GW observation provides only the tidal Love number while the moment of inertia needs to be measured through a separate electromagnetic observation of a different NS.
However, if an NS is rapidly spinning in a coalescing BNS, then it is possible to also measure $I$ directly from the GW observation in the same system where $\lambda$ is measured. In particular, the PN dynamics provides the total NS spin $\vect{S}_{1} + \vect{S}_{1, {\rm mode}}$. Meanwhile, the mode resonance depends on the background spin $\Omega_1=S_1/I_1$ (e.g., Eqs. \ref{eq:b_a_eq_EOB_0} and  \ref{eq:Delta_a_EOB}). Comparing the two measurements then directly reveals $I_1$ of the NS together with its deformability. However, it requires a high level of accuracy in the analytical templates used for detection. For example, not accounting for the tidal spin can easily bias the inferred value of $\vect{S}_1$, while nonlinear hydrodynamics studied in \cite{Yu:23a} may be confused with a change in $\Omega_1$ in the Lorentzian. The systematics can potentially lead to both offsets in the NS EoS~\cite{Pratten:22} and artificial violation of GR~\cite{Gupta:24}. A thorough exploration of testing GR with the dynamical tide would be worthy of future investigations.

\begin{acknowledgments}
We thank Nevin N. Weinberg and Phil Arras for useful discussions during the conceptualization and preparation of this work, and Marcus Haberland, Jan Steinhoff, and Alessandra Buonanno for sharing a direct comparison between our model and NR simulations. 
This work is supported by NSF grant No. PHY-2308415 and Montana NASA EPSCoR
Research Infrastructure Development under award No. 80NSSC22M0042. 
\end{acknowledgments}

\appendix
\section{Canonical vs. physical spins}
\label{appx:can_vs_phy_spin}

Defining spin carried by the perturbed fluid can be subtle even in Newtonian physics. In the main text, we adopt a \emph{canonical} definition following \cite{Friedman:78}. Because of axial symmetry of an isolated star, a transformation $\vect{\xi} \to \vect{\xi}-\delta \phi \text{\it \pounds}_\phi \vect{\xi}$ does not change the action, where the Lie derivative evaluates as $\text{\it \pounds}_\phi\xi^i = \partial_\phi \xi^i$ in a coordinate system with $(\partial_r, \partial_\theta, \partial_\phi)$ as the basis. Therefore, Noether's theorem leads to the canonical spin carried by the perturbation 
\begin{align}
    S_{z, {\rm mode}} &= \left\la -  \text{\it \pounds}_\phi \vect{\xi}, \frac{\partial \mathcal{L}_{\rm t, ns}}{\partial \dot{\vect{\xi}}} \right\ra 
    =\sum_a^+ m_a q_a q_a^\ast \epsilon_a =\sum_a^+ \frac{m_a}{\omega_a} H_a = \sum_a^+ S_a. 
\end{align}
As shown in Eq. (\ref{eq:Sij}), this is also the spin used in the analysis of \cite{Steinhoff:21}. In this appendix, we will drop the subscript ``1'' labeling its association with the first NS as we consider here only a single NS. 
The only exception is $\vect{x}_1$, the displacement of a fluid element, where the subscript is kept to distinguish it from the PN parameter $x$.
We verify that differentiating the canonical spin leads to the correct tidal torque. 
For example, from eq. (6.16) of \cite{Lai:94c}
\begin{align}
    \dot{S}_{z, {\rm mode}} 
    &=\sum_a q_a \int d^3 x_1 \nabla\cdot (\rho \vect{\xi}_a) \vect{e}_z \cdot (\vect{x}_1\times \nabla U)
    =\sum_a q_a \int d^3 x_1 \nabla\cdot (\rho \vect{\xi}_a) \frac{\partial }{\partial \phi} U^\ast, \nonumber \\
    &=\sum_a (i m_a q_a^\ast)\int d^3x_1 \rho \vect{\xi}_a^\ast\cdot (-\nabla U) = \sum_a^+ 2 m_a \omega_{a0} \epsilon_a {\rm Im}[b_a]v_a,
    \label{eq:torque_lai_94}
\end{align}
where we have integrated by parts in the first equality in the second line. This result matches our Eq. (\ref{eq:dS1z_mode_dt}). 

However, if we start from the total spin in the $z$ direction $S^z = \int d^3 x_1 \rho v_i \phi^i$ (where $v^i$ is the component of the velocity vector) 
and consider the Lagrangian perturbation of it, we have \cite{Friedman:78}
\begin{align}
    \Delta S^z = \int d^3x_1 \rho \Delta (v_i \phi^i) = \int d^3 x_1 \rho [(v_i \Delta \phi^i + \phi^i \Delta v_i) + \Delta v_i \Delta \phi^i].  
\end{align}
It can be shown that the last term leads to the canonical spin and is $\propto \xi^2$. However, at the same order, $\Delta v_i$ does not vanish and therefore also contributes to the spin at the quadratic order,
\begin{align}
    \Delta S_z^{\rm (2)}[\xi, \xi] &= S_{z, {\rm mode}} + \int d^3 x_1 \rho \phi^i \Delta v_i^{(2)}
    = \left\la\uvect{e}_z \times \vect{\xi}, \dvect{\xi} + \vect{\Omega} \times \vect{\xi} \right\ra,  \nonumber \\
    &= \sum_a^+ \left[2 q_a^\ast q_a \omega_a \la\vect{\xi}_a,  i \uvect{e}_z \times \vect{\xi}_a\ra 
    + \sum_b 2 q_a^\ast q_b \Omega \left\la \uvect{e}_z\times \vect{\xi}_a, \uvect{e}_z \times \vect{\xi}_b  \right\ra
    \right]\nonumber, \\
    &\simeq \sum_a^+  \left[\frac{\omega_a}{\omega_{a0}} \frac{S_{a}}{l} + \sum_b 2 q_a^\ast q_b \epsilon_a\frac{\Omega}{\omega_{a0}} \right].
    \label{eq:S_z_phy}
\end{align}
The second line matches eq. (K41) in \cite{Schenk:02}. We also follow the notation of \cite{Schenk:02} and use $[\xi, \xi]$ to denote the term is quadratic in $\xi$. The summation over mode $b$ runs over all modes including both signs of frequency. 
The last equality assumes incompressible fluid with $\xi_r \simeq l \xi_h$, which is a good approximation for the f-mode. 
Note that for $\Omega\simeq0$, the physical spin $\Delta S_z^{\rm (2)}[\xi, \xi]$ is about half of the canonical one. 
When $\Omega \neq 0$, the $(m_a, \omega_a)$ f-mode can couple with $(m_b=m_a, \omega_b = \pm \omega_a)$ f-modes. 

Note that while the physical spin $\Delta S_z^{\rm (2)}[\xi, \xi]$ differs from the canonical one $S_{z, {\rm mode}}$, it does not mean the tidal torques they lead to are different. This is because the linear in $\xi$ piece (eq. K8 of \cite{Schenk:02}), 
\begin{equation}
    \Delta S_z^{(1)} [\xi] \equiv \la \uvect{e}_z \times \vect{x}_1, \dvect{\xi} + 2 \vect{\Omega}\times \vect{\xi} \ra, 
\end{equation}
has a non-zero time derivative. This is further because when evaluating the time derivative of $\Delta S^{(1)} [\xi]$, the result needs to be computed to quadratic order in $\xi$ (with $a_{\rm ext}\propto \xi$). 
\begin{align}
    \Delta \dot{S}_z^{\rm (1)} &= \Delta \dot{S}_z^{\rm (1)}[\xi] + \Delta \dot{S}_z^{\rm (1)}[\xi, \xi] \nonumber \\
    &=\la \uvect{e}_z \times \vect{x}_1, \vect{a}_{\rm ext} \ra + \la \uvect{e}_z \times \vect{x}_1, (\vect{\xi}\cdot \nabla)\vect{a}_{\rm ext} \ra.
    \label{eq:bg_spin_evol}
\end{align}
The total AM transferred is 
\begin{align}
    \dot{S}_z &= \dot{S}_z^{\rm (phy, 1)}[\xi] + \dot{S}_z^{\rm (phy, 1)}[\xi, \xi] + \dot{S}_z^{\rm (phy, 2)}[\xi, \xi] \nonumber \\
    &= \la \uvect{e}_z \times \vect{x}_1, \vect{a}_{\rm ext} \ra + \la \uvect{e}_z \times \vect{x}_1, (\vect{\xi}\cdot \nabla)\vect{a}_{\rm ext} \ra + \la\uvect{e}_z \times \vect{\xi}, \vect{a}_{\rm ext} \ra \nonumber \\
    &=\int d^3 x_1 \rho \left[\varepsilon_{zki}\xi^k a_{\rm ext}^i + \varepsilon_{zij}x_1^i \xi^k \nabla_k a_{{\rm ext}}^i \right] \nonumber \\
    &=\int d^3 x_1 \rho \left[\xi^i \nabla_i (\varepsilon_{zkj} x_1^k a_{\rm ext}^j)\right] =\int d^3 x_1 \rho \left\{\vect{\xi} \cdot \nabla [\uvect{e}_z \cdot (\vect{x}_1 \times \vect{a}_{\rm ext}) ]  \right\} \nonumber \\
    &=\int d^3 x_1 \nabla \cdot (\rho \vect{\xi}) [\uvect{e}_z \cdot (\vect{x}_1\times \nabla U)]. 
\end{align}
where, from the second to the third line, we have dropped the $\dot{S}_z^{\rm (phy, 1)}[\xi]$ contribution as it vanishes for electric type modes like the f-modes; the last two lines respectively match eq. (K52) of \cite{Schenk:02} and eq. (6.16) of \cite{Lai:94c}, and they match the torque computed from the canonical value as shown in Eq. (\ref{eq:torque_lai_94}). 

As the torque matches, the total spins computed by the two ways are the same up to an irrelevant constant, which can be further set to zero by requiring both spins to vanish as $\xi\to0$,
\begin{equation}
    I \Omega[\xi, \xi]  + \Delta S^{(2)}_z[\xi, \xi] = S_{z, {\rm mode}}[\xi, \xi]. 
\end{equation}
In other words, when the physical spin is used, one needs to account for not only the $S^{(\rm phy, 2)}_z[\xi, \xi]$ piece, but also the evolution of the background spin to quadratic order (i.e., $\Omega[\xi, \xi]$ from eq. \ref{eq:bg_spin_evol}). However, the background spin is viewed as a constant while all the AM is carried by the mode in the canonical picture. As the latter is more convenient to track in our modal expansion framework, we adopt the canonical description in the main text.

\section{Mode amplitudes and NS multiples}
\label{appx:q_a_to_multipoles}

In Eq. (\ref{eq:Qij}) we present the relation between the (inertial frame) mode amplitude $q_a$ and the mass quadrupole moment $Q^{ij}$ assuming only the $l=2$ f-modes. A more general connection for arbitrary $l$, including all eigenmodes entering the expansion of Eq. (\ref{eq:mode_decomp}), and allowing for generic normalization conditions is presented in this appendix for completeness. 

We start from the mass multipole moments, following the definition in \cite{Poisson:14}, as (see also \cite{Yu:23a})
\begin{align}
    Q_{lm} &\equiv \int d^3 x_1 \rho \vect{\xi} \cdot \nabla(x_1^l Y_{lm}^\ast) = \int d^3 x_1 \rho \sum_a q_a \vect{\xi}_a^\ast \cdot \nabla(x^l_1 Y_{lm}) = M_1 R_1^l \sum_a I_a q_a \nonumber \\
    &= \frac{M_2}{r^{l+1}} R_1^{2l+1} W_{lm} e^{-im\phi} \sum_a \left(\frac{M_1^2 / R_1}{\omega_{a0} \epsilon_a}\right) I_a^2 \frac{b_a}{v_a},
\end{align}
where we have used the definition of the overlap integral $I_a$ in Eq. (\ref{eq:overlap}) and the summation runs over modes with $s_a=\pm$. In the last line, we further normalize each mode's amplitude by its tidal drive, Eq. (\ref{eq:v_a}). 
The related STF tensor is 
\begin{align}
    Q^{\la L\ra} = N_l \sum_m \mathcal{Y}_{lm}^{\la L\ra \ast} Q_{lm},
\end{align}
where $N_l=4\pi l!/(2l+1)!!$ and $N_l \mathcal{Y}_{lm}^{\la L \ra} \mathcal{Y}_{lm',\la L \ra}^\ast = \delta_{m m'}$. 

Meanwhile, we want to retain the form for each $(l, m)$ harmonic that
\begin{equation}
    Q_{lm} = - \lambda_{lm} \mathcal{E}_{lm},
\end{equation}
with $\lambda_{lm}$ the deformability. 
The tidal potential is, 
\begin{align}
    \mathcal{E}_{\la L \ra} &= N_l \sum_m \mathcal{Y}_{lm}^{\la L\ra \ast} \mathcal{E}_{lm},\nonumber \\
    &= - M_2 \partial_{\la L \ra_{\vect{x}_1} }\frac{1}{|\vect{x}_1 - \vect{r}|}\Bigg{|}_{\vect{x}_1=0} 
    = (-1)^{l+1} M_2\partial_{\la L \ra_{\vect{r}} }\frac{1}{|\vect{x}_1 - \vect{r}|}\Bigg{|}_{\vect{x}_1=0} \nonumber \\
    &= -(2l-1)!! M_2 \frac{n_{\la L\ra}}{r^{l+1}} = - (2l-1)!!N_l \frac{M_2}{r^{l+1}} \sum_m \mathcal{Y}_{lm}^{\la L\ra \ast} Y_{lm}^\ast,
\end{align} 
and $\vect{n} = \vect{r}/r$ is the companion's location in a frame centered on the center of $M_1$ with $\vect{x}_1 = 0$.  
This leads to
\begin{align}
    \frac{\lambda_{lm}}{R_1^{2l+1}}&= \frac{4\pi}{(2l+1)!!} \sum_{m_a=m}^{s_a=\pm} \left(\frac{M_1^2/R_1}{\omega_{a0}\epsilon_a}\right)I_a^2  \frac{b_a}{v_a}, \nonumber \\
    &\equiv \frac{\lambda_{l}}{R_1^{2l+1}} \kappa_{lm}=\frac{2}{(2l-1)!!} k_{l} \kappa_{lm}, 
\end{align}
where
\begin{align}
    \frac{\lambda_{l}}{R_1^{2l+1}} = \frac{2}{(2l-1)!!}k_{l} =  \frac{4 \pi}{(2l+1)!!}\sum_{m_a=0}^{s_a=\pm}  \left(\frac{M_1^2/R_1}{\omega_{a0}\epsilon_a}\right) I_a^2
\end{align}
is the adiabatic deformability and $k_l$ the adiabatic Love number, and  
\begin{align}
    \kappa_{lm} = \frac{\sum_{m_a=m}^{s_a=\pm} \left(\frac{M_1^2/R_1}{\omega_{a0}\epsilon_a}\right)I_a^2 \left(\frac{b_a}{v_a}\right)}{\sum_{m_a=0}^{s_a=\pm} \left(\frac{M_1^2/R_1}{\omega_{a0}\epsilon_a}\right)I_a^2 },
\end{align}
the effective Love number. 

In terms of $\lambda_l$ and $\kappa_{lm}$, the interaction energy at a particular $l$ is conveniently given 
\begin{equation}
    H_{\rm int, l}=\sum_m H_{\rm int, lm} = -\frac{(2l+1)!!}{4\pi} \left(\frac{M_2}{r^{l+1}}\right)^2 \lambda_l \sum_m W_{lm}^2 \kappa_{lm}. 
\end{equation}

\bibliography{ref}

\end{document}